\documentclass[sigconf,screen]{acmart}
\usepackage{xspace}
\newcommand{\methodbase}{CureLLM\xspace} 
\usepackage{amsmath}
\usepackage{amsthm}
\usepackage{supertabular}
\usepackage{booktabs}
\usepackage{xcolor}
\usepackage{array} 
\usepackage{booktabs}
\AtBeginDocument{%
  }
\renewcommand\footnotetextcopyrightpermission[1]{} 
\acmConference[MM '26]{ACM Multimedia 2026}{November 10--14, 2026}{Rio de Janeiro, Brazil}
\begin{document}

\title{Edge-Aware Curvature Modeling for Graph Understanding in Large Language Models}


\author{Zhenghong Lin}
\affiliation{%
  \institution{College of Computing and Data Science, Nanyang Technological University}
\institution{Computational Intelligence Lab}
  \country{Singapore}}
\email{hongzhenglin970323@gmail.com}

\author{Zhibin Shi}
\affiliation{%
  \institution{College of Computer and Data Science, Fuzhou University}
 \institution{Fujian Provincial Key Laboratory of Network Computing and Intelligent Information Processing}
   \city{Fuzhou}
  \country{China}
}
\email{zbshi0593@163.com}

\author{Hongyang Dong}
\affiliation{%
  \institution{College of Computer and Data Science, Fuzhou University}
 \institution{Fujian Provincial Key Laboratory of Network Computing and Intelligent Information Processing}
   \city{Fuzhou}
  \country{China}
}
 \email{Donghy0305@163.com}

\author{Xinjie Ye}
\affiliation{%
  \institution{Department of Statistics, University of Wisconsin-Madison}
  \country{USA}
 }
 \affiliation{
  \institution{Newland Digital Technology Co., Ltd.}
  \city{Fuzhou}
  \country{China}
 }
\email{yexj@newland.com.cn}

\author{Yuhong Chen}
\affiliation{%
  \institution{Key Laboratory of Multimedia Trusted Perception and Efficient Computing, Ministry of Education of China, Xiamen University}
  \city{XiaMen}
  \country{China}}
\email{yhchen2320@163.com}

\author{Shiping Wang}
\affiliation{%
  \institution{College of Computer and Data Science, Fuzhou University}
 \institution{Fujian Provincial Key Laboratory of Network Computing and Intelligent Information Processing}
   \city{Fuzhou}
  \country{China}
}
\email{shipingwangphd@163.com}


\renewcommand{\shortauthors}{Trovato et al.}

\begin{abstract}
  Recently, graph-aware Large Language Models (LLMs) have shown promising capabilities in jointly modeling graph-structured data and textual information. Existing approaches typically employ a graph encoder and a frozen LLM to obtain node representations from graph and textual views, followed by node-level alignment to bridge the two modalities. 
However, such alignment mechanisms primarily focus on node information while overlooking edge-level structures, leading to suboptimal information propagation across views.  In this work, we conduct a comprehensive theoretical analysis to uncover why node-level alignment is insufficient for aligning textual and graph
representations. Specifically, we prove theoretically for the first time that (1) neglecting edge information leads to suboptimal solutions and (2) negatively curved edges induce bottlenecked information flow, giving rise to the over-squashing phenomenon between graph and textual views. To address the two challenges, we innovatively proposed a \methodbase framework of Curvature-enhanced Graph Representations for Large Language Model whose goal is to inject the signals of edge information into the existing LLMs. Specifically, \methodbase first introduces the training-free textual  prompt mechanism to make the LLM model generate the output directly based on the edge-aware prompt without learnable parameter costs. Furthermore, a novel curvature-aware graph representation learning is designed to capture the edge structure information to enhance the downstream tasks, where the message passing between text and graph representations only depends on edges with positive curvature. Finally, we conduct evaluations with 20 different compared methods on 11 real-world datasets from various domains and the experiment results demonstrate the superiority of our proposed \methodbase framework.
\end{abstract}




\maketitle

\section{Introduction}
 
\begin{figure*}[!t]
    \centering
\includegraphics[width=1\linewidth]{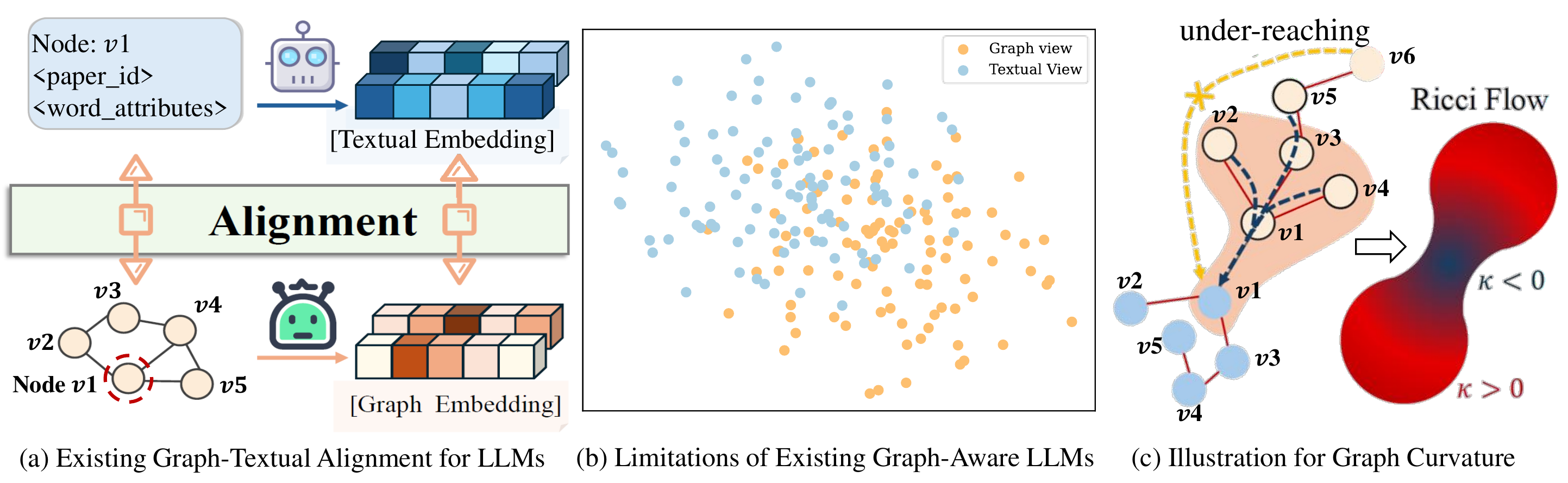}
    \caption{A toy example of large language models (LLMs) with graph understanding.
    (a) The existing paradigm of graph-textual LLMs;
    (b) The limitations of existing graph-textual alignment mechanisms in LLMs;
    (c) The illustration for graph curvature.
    }
    \label{toy2}
\end{figure*}

 With the rapid growth of online data scale \cite{luo2024chatkbqa, DBLP:conf/nips/DuLJLN24}, the large language models (LLMs) have achieved great success in many domains \cite{ren2024survey, hu2024uncertainty, NEURIPS2024_305b2288}. 
In order to transfer the rich knowledge of LLMs to the practical tasks based on the graph-structured data \cite{lin2023automatic}, such as recommender systems \cite{lin2024enhancing}, social networks \cite{DBLP:conf/kdd/MengWHL0Z24} and disease prediction \cite{DBLP:conf/www/0002TZ025},  graph-aware LLMs have become a critical research area in recent years \cite{DBLP:conf/nips/0001WZCJ0C024, wang2023can}.
To enable LLMs to recognize graph structure data, most previous research \cite{tang2024higpt} has explored
projecting the
textual attributes and graph topology structure into the same latent space to enhance the 
 representation learning in LLMs. Specifically, as illustrated in Fig. \ref{toy2}(a), the existing graph-textual paradigm of LLMs may use the frozen LLM encoder and trainable graph adapter to extract the textual and graph embeddings for the node $v_1$. Then, some alignment mechanism, such as supervised learning \cite{chen2023agnn} or contrastive learning \cite{wei2024llmrec}, is adopted to align the node embeddings under textual and graph views for node $v_1$.

However, as a relatively nascent research area, recent studies and empirical observations suggest that such graph-textual alignment mechanism often leads to the suboptimal performance\cite{zheng2025cross}. 
To gain initial insights into this issue, we visualize the learned representations of existing graph-aware LLMs\cite{DBLP:conf/sigir/Tang00SSCY024}, as shown in Fig.~\ref{toy2}(b). We observe that the alignment between textual and graph representations is often partially aligned, especially for samples near decision boundaries. 
This gap leaves a fundamental question unanswered: \textbf{Why is node-level alignment insufficient for aligning textual and graph representations?}

%
%
To explain the above question, we  attempt to analyze the existing paradigm of graph-aware LLMs from a theoretical perspective. Our findings indicate that the limitations of node-level alignment stem from insufficient modeling of information propagation over graph structures, which can be attributed to two main reasons. 

\noindent\textbf{(1) Structural Constraints from Edge-Level Information.}
Beyond node-level representations, information propagation is inherently governed by edge-level structures, which define the pathways of message passing in graphs. However, existing graph-aware LLMs largely overlook the role of edges in controlling how information flows across the graph and node-level alignment implicitly treats all connections uniformly. In particular, different edges may exhibit varying capacities for information transmission, leading to heterogeneous propagation patterns that cannot be captured by node-only alignment mechanisms. We further analyze the limitations of existing graph-textual alignment mechanisms in LLMs, as formalized in \textbf{Theorem 1} (Section~\ref{motivation_theorem}).

\noindent\textbf{(2) Information Compression in Node-Level Alignment.}
In existing methods, alignment is established by matching node embeddings across textual and graph views. However, each node representation must aggregate information from its multi-hop neighborhood. Take Fig. \ref{toy2}(c) as an example, when alignment is established solely through a shared node (i.e., node $v_1$ in the textual view aligned with $v_1$ in the graph view), messages from multiple nodes (e.g., $v_2, v_3, v_4, v_5,$ and $v_6$) are forced to propagate through the same alignment bottleneck at $v_1$. This many-to-one interaction induces severe path-level congestion, where structurally diverse information is compressed into a single-node representation. 
The above problem, which compresses an exponentially growing amount of information into fixed-size node features and hinders message passing between nodes, is referred to as the \emph{over-squashing} phenomenon in previous works \cite{DBLP:conf/icml/AttaliBP24, attali2024delaunay}. This phenomenon is further theoretically characterized in \textbf{Theorem 2} (Section~\ref{motivation_theorem}), which formalizes the impact of such bottlenecked information propagation.

To address the above two challenges, we innovatively propose a \methodbase framework of {\bf Cur}vature-{\bf e}nhanced Graph Representations for {\bf L}arge {\bf L}anguage {\bf M}odel, consisting of the training-free textual prompt mechanism and curvature-aware graph representation learning. Specifically, our \methodbase framework aims to explicitly model edge-level information flow while mitigating the information compression issue in node-level alignment. 
For structural constraints from edge-level information, we design a training-free textual prompting mechanism to inject the edge information into pre-trained LLMs without any additional task-specific training, which can reduce the cost of training. 
For information compression in node-level alignment, we propose a curvature-aware graph representation learning module to leverage curvature as a geometric tool to characterize information propagation on graphs.
As established in prior works \cite{DBLP:conf/icml/FarzamTS24}, Ricci curvature provides a principled way to characterize such propagation behavior: negatively curved regions correspond to bottlenecked information flow, while positively curved regions indicate smoother information transmission, as shown in Fig. \ref{toy2}(c).
Based on this observation, a two-step curvature alignment enhancement is exploited to alleviate the information bottleneck between the graph and textual views. Specifically, we enhance information propagation for inter-view alignment by promoting interactions along edges with positive curvature, while adopting graph convolution for intra-view representation learning to ensure effective transfer of structural information across views. 
Finally, after the alignment mechanism, the obtained final textual-graph
fusion representations are fed to the different heads for various downstream tasks, including node classification, link prediction and graph question answering.

The contributions of this paper are outlined as follows:
 
 \begin{itemize}

    \item \textit{(1) About Novelty: }To the
best of our knowledge, our work is the first to introduce the curvature into the graph-aware LLMs to alleviate the over-squashing phenomenon between texutal and graph views.
    \item
    \textit{(2) About Effective Model Design: } Our \methodbase is the first framework to exploit the edge structure information to enhance the alignment between textual and graph representations in graph-aware LLMs.

    \item \textit{(3) About Theoretical Analysis: } We prove theoretically for the first time that the lack of edge information
can lead to a suboptimal solution, and the edges with negative curvature are the main reason for the information bottleneck between the graph and textual views. 

\item \textit{(4) About Extensive Experiments: } We conduct comprehensive experimental evaluations for three graph tasks, and experimental results on $11$ real-world public datasets with $20$ compared methods show the superiority of our \methodbase. 
    
\end{itemize}

\section{Theoretical Motivation and Preliminary}
\label{motivation_theorem}

In this section, we first provide a theoretical analysis of the error upper bound and the message passing mechanism in graph representation learning with LLMs, in order to better understand the limitations of existing graph-aware LLM paradigms.
Based on this analysis, we derive two key motivations for our work:
(1) the existing paradigm for graph LLMs using only node alignment while ignoring the structural information of edges can lead to suboptimal solutions and (2) negatively curved edges are
those causing the graph bottleneck and thus leading to the partially indistinguishable phenomenon. The detailed theoretical derivations are provided in \textbf{Appendix A}. Finally, we introduce the necessary preliminaries for our method, including fundamental concepts of curvature and the background is presented in \textbf{Appendix B}.

\noindent{\bf Theoretical Motivation.} Firstly, we analyze the Generalization Error Upper Bound~\cite{DBLP:conf/icml/CaoXDZH24} of graph representation learning for large language models. The smaller the upper bound of the
generalization error, the better generalization ability for the model. Given the graph $\mathcal{G} = (\mathcal{X}, \mathcal{Y}, \mathcal{E})$, we define the training set $\mathcal{D} = \{(\boldsymbol{x}_i,\boldsymbol{y}_i)\}_{i=1}^N  \in \mathcal{X} \times \mathcal{Y}$, where $\mathcal{X}, \mathcal{Y}, \mathcal{E}$ are the node, label and edge sets. Then, we denote
the graph projection function $f(\boldsymbol{x}): \mathcal{X} \rightarrow \mathcal{Y}$ and $f(\boldsymbol{x})$ is related to the information of the node itself and the aggregation mechanism through the edges. Therefore, the function $f(\boldsymbol{x})$ can be defined as $f(\boldsymbol{x}) = \omega_n f_n(\boldsymbol{x}_n) + \omega_e f_e(\boldsymbol{x}_e)$, where $\omega_n$ and $\omega_e$ are the confidence weights for nodes and edges respectively. When defining $\ell$ as a convex logistic loss function, we can obtain the following theorem and the
full theoretical proof of {\bf Theorem 1 } is given in {\bf Appendix A.1}. 

\noindent
{\bf Theorem 1. }(Error Upper Bound for Graph Representation Learning in LLMs) \emph{Let $\hat{e}(f_n)$ and $\hat{e}(f_e)$ denote the empirical errors
of nodes and edges in the graph, and $\mathcal{H}$ is
hypothesis set, where $f \in H$, and 
$\mathcal{R}_N(\mathcal{H})$ is the Rademacher complexities~\cite{trappenberg2018fundamentals}. With a confidence level of $1-\Delta$
$(0 < \Delta < 1)$
, we have generalization error (GE):
}

\begin{equation}
\begin{aligned}
GE(f) & \leq  2\mathcal{R}_N(\mathcal{H}) + \sqrt{ \frac{2\ln(1/\Delta)}{N} } 
+ \hat{{e}}(f_n) + \hat{{e}}(f_e)
+ \\
&\frac{1}{2}\Bigg[ \underbrace{Cov(\omega_n, \ell_n)}_{\text{Node Covariance}} +
\underbrace{ Cov(\omega_e, \ell_e)}_{\text{Edge Covariance}} \underbrace{- Cov(\omega_n, \ell_e) - Cov(\omega_e, \ell_n)}_{{\text{Node-Edge Covariance}}}  \Bigg],
\end{aligned}
\end{equation}
where $Cov(\omega, \ell)$ is the covariance between the confidence weight $\omega$ and the loss function $\ell$. Because the empirical errors $\hat{{e}}(f_n), \hat{{e}}(f_e)$ and Rademacher complexities $\mathcal{R}_N(\mathcal{H})$ are constant, the lower $GE(f)$ bound is achieved when $Cov(\omega_n, \ell_n) \textless 0$, $Cov(\omega_e, \ell_e) \textless 0$ and $Cov(\omega_n, \ell_e) \textgreater 0$, $Cov(\omega_e, \ell_n) \textgreater 0$ are satisfied.
{\bf Theorem 1} shows not only that the generalization ability of the model also depends on the edge structure, but it also reveals how edge information should vary across nodes. Specifically, node and edge covariances ($Cov(\omega_n, \ell_n) \textless 0$ and $Cov(\omega_e, \ell_e) \textless 0$) imply that when the node or edge embeddings are trained well enough for message passing, we should increase the weights of these nodes and edges. On the contrary, the node-edge covariance ($Cov(\omega_n, \ell_e) \textgreater 0$ and $Cov(\omega_e, \ell_n) \textgreater 0$) suggests that when node information is insufficient for graph propagation, the weights of edges need to be increased to alleviate the above phenomenon.

Secondly, we analyze the existing message passing paradigm for the alignment between the graph and text contents of LLMs and try to explain from a geometric perspective that edges with negative curvature limit the potential for graph representation learning in LLMs.  For the convenience of explanation, we introduce the definitions of the message passing mechanism \cite{wu2023interpretable}.

\noindent
{\bf Definition 1.} (Graph Message Passing Mechanism) \emph{Given the adjacency matrix for the graph $\mathcal{G} = (\mathcal{X}, \mathcal{Y}, \mathcal{E})$ with $d$-dimensional representations, we denote the hidden representation of node $i$ at layer $l$ as $\boldsymbol{h}_i^{(l)}$, where $\boldsymbol{h}_i^{(0)} = \boldsymbol{x}_i$.
Assume that message
function is $\psi_l$ and the update function is $\phi_l$, the graph message passing mechanism can be reformulated as}
\begin{equation}
\begin{aligned}
\boldsymbol{h}_i^{(l+1)} = \phi_l ( \boldsymbol{h}_i^{(l)}, \sum_{j=1}^{n} \boldsymbol{A}_{ij} \psi_l ( \boldsymbol{h}_i^{(l)}, \boldsymbol{h}_j^{(l)} ) ).
\label{graph_message_passing}
\end{aligned}
\end{equation}
Then, we introduce the definition of curvature \cite{li2022curvature} to further analyze the graph mechanism. Following the previous works about the Ollivier-Ricci curvature~\cite{ye2019curvature}, the curvature of the edge between nodes $i$ and $j$ can be written as

\noindent
{\bf Definition 2.} (Graph Ollivier-Ricci Curvature) \emph{Given the the graph $\mathcal{G} = (\mathcal{X}, \mathcal{Y}, \mathcal{E})$, we denote the set of neighboring nodes of a node $i \in \mathcal{X}$ as $\mathcal{N}(i)$ and 
a probability measure $\boldsymbol{m}_x^\alpha$ with
a parameter $\alpha$ within $[ 0, 1 ]$. For any edge between node $i$ and $j$, 
the coarse Ricci curvature $\boldsymbol{\kappa}$ on edge
$(i, j)$ is defined by comparing the Wasserstein distance $W(\boldsymbol{m}_i, \boldsymbol{m}_j)$ to the distance d(i, j):}
\begin{equation}
\begin{aligned}
\boldsymbol{\kappa}_{ij} = 1 - \frac{W(\boldsymbol{m}_i, \boldsymbol{m}_j)}{d(i, j)} 
 \quad \text{s.t.} \quad 
  \boldsymbol{m}_x^{\alpha}(i) = 
\begin{cases}
\alpha & \text{if } x = i \\
\frac{1 - \alpha}{k} & \text{if } i \in \mathcal{N}(x) \\
0 & \text{otherwise}
\end{cases}.
\label{ricii_curvature}
\end{aligned}
\end{equation}
After defining the $r$-layer set $S_r(i) = \{ j\in \mathcal{X}: d_G(i,j) = r\}$ and degrees of nodes $i, j$ as $d_i, d_j$, where $d_G$ is the standard shortest-path distance on the graph and $r \in \mathbb{N}$ is the receptive field in graph message passing, we can obtain {\bf Theorem 2} and  the full proof is given in {\bf Appendix A.2}.

{\bf Theorem 2.} (Negative Curvature in Graph Representation Learning with LLMs) \textit{Consider message passing and curvature as in Equation \eqref{graph_message_passing} and \eqref{ricii_curvature}. Let node $i$ under the graph view and node $j$ under the textual view with $d_i \leq d_j$. Assume that the derivatives of functions under graph-graph and text-text message passing are bounded with $|\nabla \phi_l| \leq \alpha$ \textit{and} $|\nabla \psi_l| \leq \beta$, and the message passing functions between graph-text nodes are bounded with $|\nabla \phi_l^{\prime}| \leq \alpha^{\prime}$ \textit{and} $|\nabla \psi_l^{\prime}| \leq \beta^{\prime}$, \textit{for each} $0 \leq l \leq L-1$, \textit{with} the depth of message passing $L \geq 2$. Following \cite{DBLP:conf/iclr/ToppingGC0B22}, \textit{there exists $\delta$ and $\gamma_{\text{max}}^{-1}$ such that } $0 < \delta < \left(\max\{d_i, d_j\}\right)^{-\frac{1}{2}},\ \delta < \gamma_{\text{max}}^{-1}$ \textit{and } $\boldsymbol{\kappa}_{ij} \leq -2 + \delta$.}
\textit{Then given $Q_j \subseteq S_2(i)$ satisfying $|Q_j| > \delta^{-1}$ and for $0 \leq l_0 \leq L - 2$, we have}
\begin{equation}
\begin{aligned}
\frac{1}{|Q_j|} \sum_{k \in Q_j} \left| \frac{\partial \boldsymbol{h}_k^{(l_0 + 2)}}{\partial \boldsymbol{h}_i^{(l_0)}} \right| < (\alpha \beta) ( \alpha^{\prime} \beta^{\prime}) \delta^{\frac{1}{4}}.
\label{negative_ricii_curvature}
\end{aligned}
\end{equation}
The interpretability of {\bf Theorem 2} is that if we
have a negative curvature for $edge(i,j)$ ($\boldsymbol{\kappa}_{ij} \leq -2 + \delta$), there may exist a large number of nodes $k$ ($k \in Q_j$), where the message passing from node $i$ under the graph view to nodes $k$ under textual view is affected, even though the distance is $2$ for these nodes $k$ to node $i$ (from layers $l_0$ to $l_0 + 2$).

\noindent{\bf Preliminary for Prompt Learning.} Prompt learning \cite{DBLP:conf/nips/Castillo-Bolado24} is an effective fine-tuning strategy to advance the generalization from LLMs to a specific downstream task \cite{sun2024prompt}. Specifically, these prompt-based approaches \cite{DBLP:conf/nips/0001L00024, DBLP:conf/nips/0001MZ0GP24} usually add a small
number of additional embeddings (e.g., the explanation why the model makes the prediction or a detailed description for the task) along with model inputs, while  the LLM
models are kept frozen. In this paper, our prompts are textual forms including additional information or constraints to guide behaviors of the model, such as a single sentence for the title of a node or a longer passage for the description of a specific task. Formally, the training process of LLMs can be defined to optimize a conditional probability distribution $p(\boldsymbol{y}|\boldsymbol{s})$, where $\boldsymbol{s} = (\boldsymbol{s}_1, \dots, \boldsymbol{s}_{L_I})$ is  a textual sequence of tokens for model inputs and $L_I$ is the length of the input sentences. After introducing the prompt $\boldsymbol{s}^p = (\boldsymbol{s}^p_1, \dots, \boldsymbol{s}^p_{L_P})$, the new paradigm of LLMs can be reformulated to compute the conditional probability distribution $p(\boldsymbol{y}|\boldsymbol{\hat{s}}) = \prod_{i=1}^{m} p(\boldsymbol{y}_i \mid \boldsymbol{y}_{<i}, \boldsymbol{\hat{s}})$. Here $\boldsymbol{\hat{s}} = (\boldsymbol{s}^p_1, \dots, \boldsymbol{s}^p_{L_P}, \boldsymbol{s}_1, \dots, \boldsymbol{s}_{L_I})$ is the token sequence obtained by concatenating the textual prompt $\boldsymbol{s}^p$ and input $\boldsymbol{s}$. The probability $p(\boldsymbol{y}|\boldsymbol{\hat{s}}) = \prod_{i=1}^{m} p(\boldsymbol{y}_i \mid \boldsymbol{y}_{<i}, \boldsymbol{\hat{s}})$ means the
probability of generating token $\boldsymbol{y}_i$ given $\boldsymbol{y}_{<i}$ and $\boldsymbol{\hat{s}}$. 
More details about the {\bf Related Work} can be found in {\bf Appendix B.}


\begin{figure*}[t]
    \centering
\includegraphics[width=0.9\linewidth]{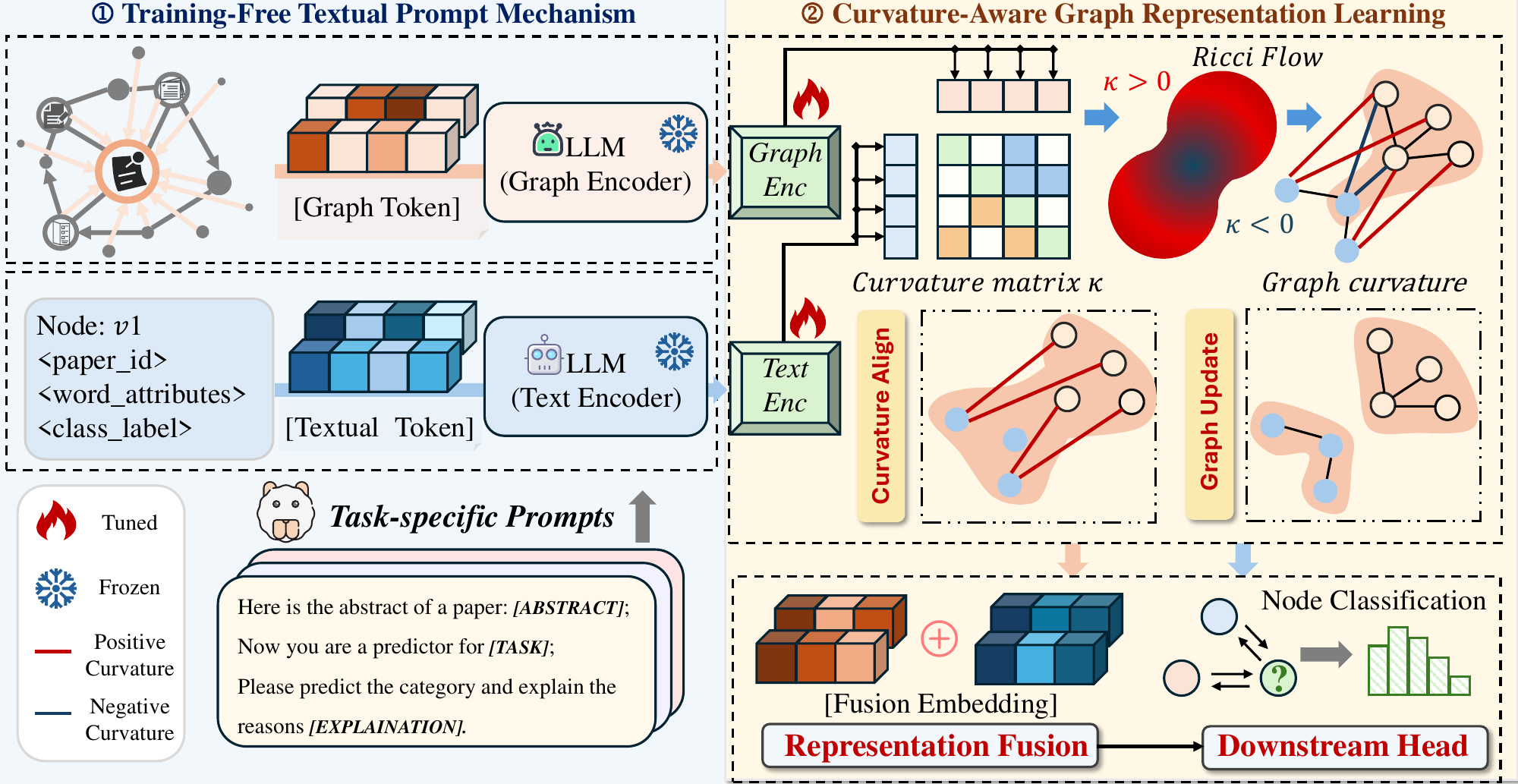}
    \caption{The overall framework of \methodbase. In the left part, a parameter-free mechanism is employed to generate the desired output without any additional task-specific training. In the right part, the  Curvature-Aware Graph Representation Learning is designed to align the embeddings between textual and graph views based on the graph curvature.
    }
    \label{fig:frame}
\end{figure*}

\section{Methodology}
\subsection{Problem Statement and Architecture Overview}

The goal of our proposed \methodbase method is to inject the signals of edge attributes into the existing learning paradigm for the graph-driven LLMs to improve the  generalization ability of LLMs for text contents with graph structures, e.g., text-attributed graphs (TAGs). Specifically, given the original node feature $\boldsymbol{X}$ of the graph and textual sequence $\boldsymbol{S} = \{\boldsymbol{S}_1, \boldsymbol{S}_2, \dots, \boldsymbol{S}_i, \dots \boldsymbol{S}_N \}$ for each node, we use the frozen LLM and graph encoder that have been well pre-trained on large-scale datasets to obtain the embeddings $\boldsymbol{X}^{txt} \in \mathbb{R}^{N \times d_t}$ and $\boldsymbol{X}^{gra} \in \mathbb{R}^{N \times d_g}$ under the textual and graph views, respectively. 
Here, $N$ is the number of nodes in the graph structure, $d_t$ and $d_g$ are the dimensions of the latent space, and $\boldsymbol{S}_i$ means the textual contents of the model input for the $i$-th node. Enhanced by the adjacency matrix $\boldsymbol{A} \in \mathbb{R}^{N \times N}$ and edge curvature $\boldsymbol{\kappa} \in \mathbb{R}^{N \times N}$, our proposed \methodbase can generate the aligned fusion embedding $\boldsymbol{X}^f \in \mathbb{R}^{N \times d}$, where the values of $\boldsymbol{A}_{ij}$ means the connection relations between nodes $i$ and $j$ and $\boldsymbol{\kappa}_{ij}$ is the value of curvature for $edge(i,j)$. Note that, after extracting the refined final embedding, we fed the fusion embedding $\boldsymbol{X}^f$ to the different heads of various downstream tasks (we focus on three main graph tasks in this paper: node classification, link prediction and graph question answering). 
In summary, the inputs of our \methodbase are the node feature $\boldsymbol{X}$, the adjacency matrix $\boldsymbol{A}$ and the textual sequence $\boldsymbol{S}$. The outputs are the refined fusion embedding $\boldsymbol{X}^f$. The overall architecture is provided in Fig. \ref{fig:frame}, which consists of the training-free textual
prompt mechanism and the curvature-aware
graph representation learning.

\vspace{-2mm}
\subsection{Training-Free Textual
Prompt Mechanism}
Instead of pre-training the LLM for downstream tasks via specific objective engineering \cite{kopf2023openassistant}, we aim to directly guide the LLM to incorporate edge structural information without additional training. Inspired by the success of prompt learning \cite{liu2023pre} with an entirely unsupervised way to solve a great
number of tasks, we propose a novel training-free textual prompt mechanism that explicitly encodes edge-aware structural signals into textual inputs without training costs. Specifically,
our prompt mechanism integrates both node-level attributes and edge-level structural information into the textual view. Given a target node, we construct edge-aware prompts by incorporating its structural context, such as neighboring nodes, connectivity patterns, or relation-aware descriptions, into the textual input. In this way, edge information is transformed into a textual form that can be naturally processed by the LLM.
We adopt two frozen encoders to extract representations under textual and graph views (textual encoder and graph encoder), respectively. Then, we initialize and freeze the two encoders using the graph-aware large language model weights which have been well pre-trained on the large corpus of text and graph data. For example, we use the frozen Llama \cite{touvron2023llama} as the textual encoder and a frozen graph encoder in \cite{tape}. 

To effectively transfer the textual and graph knowledge from general frozen weights to unknown downstream tasks, our training-free prompt ensemble the techniques for textual-prompt enhancement by constructing diverse task-specific captions, attributes, or high-level concepts of downstream tasks. The more details about designs for the prompt can be referred to {\bf Appendix A.3}. We take the node classification task as an example and 
the general template of the prompt is:
\begin{equation}
\label{prompt}
\text{
  \parbox{0.9\columnwidth}{ 
    \textit{Here is the abstract of a paper: [\textbf{ABSTRACT}]; 
    This paper belongs to the [\textbf{TOPIC}]; 
    The title of this paper is [\textbf{TITLE}]; The citation relationships of this paper are [\textbf{EDGE PROMPTS}]; [\textbf{OTHER PROMPTS}]; 
    Now you are a predictor for [\textbf{TASK}]; 
    Please predict this paper may belong to what category and explain the reasons [\textbf{EXPLANATION}].}
  }
}
\end{equation}
The above locations \textit{ [\textbf{ABSTRACT}], [\textbf{TOPIC}], [\textbf{TITLE}], [\textbf{EDGE PROMPTS}]} are the data-specific textual prompts and \textit{ [\textbf{TASK}], [\textbf{EXPLAINATION}]} are the task-specific hand-crafted prompts. The identifier \textit{[\textbf{OTHER PROMPTS}]} means that readers can optionally add relevant prompts meeting the requirements for various tasks to enhance the transferability of LLMs without any additional task-specific training. For node $i$ with the feature $\boldsymbol{X}_i$ and its textual data $\boldsymbol{\hat{S}}_i = \{ \boldsymbol{S}^P_{i,1}, \dots, \boldsymbol{S}^P_{i, L_P}, \boldsymbol{S}_{i, 1}, \dots, \boldsymbol{S}_{i, L_I} \}$, where $L_P$ and $L_I$ are the length of prompts and the original text, we can extract the low-dimensional representation by the frozen LLM and graph encoder. The above process can be described as follows:
\begin{equation}
\begin{aligned}
\boldsymbol{X}_i^{txt} = EncoderLLM(\boldsymbol{\hat{S}}_i), \quad \boldsymbol{X}_i^{gra}= EncoderGraph(\boldsymbol{X}_i).
\label{Frozen_encoder}
\end{aligned}
\end{equation}
Here, $\boldsymbol{S}^P, \boldsymbol{S}$ are the textual tokens of prompts and original inputs respectively, and $\boldsymbol{\hat{S}}$ is the concatenated textual contents. Then, the textual representation  $\boldsymbol{X}^{txt} \in \mathbb{R}^{N \times d_t}$ and graph representation $\boldsymbol{X}^{gra}\in \mathbb{R}^{N \times d_g}$ are fed to the next curvature-aware graph representation learning for fusion, where $N$ is the number of nodes and $d_t, d_g$ are the dimension of latent embeddings. 
Compared with existing prompt learning \cite{liu2024can, DBLP:conf/acl/HanCBS24}, the training-free text prompting mechanism we propose has the following advantages:
(1) Since only frozen weights are used for graph and textual encoders, our mechanism can easily ensemble existing graph-aware LLMs that have been pre-trained on other datasets; (2) Due to no need for training from scratch, our mechanism can avoid the additional computational overhead.

\subsection{Curvature-Aware
Graph Representation Learning}

After the enhancement by our training-free textual prompt mechanism, we aim to learn the refined graph-textual fusion representation for improving the performance for various downstream tasks. From the previous discussion in Section \ref{motivation_theorem}, we can learn that the existing fusion mechanism only considers the node information while ignoring the edge attributes may lead to suboptimal solutions (Refer to {\bf Theorem 1}). By discussing curvatures for edge attributes, we can know the edges with negative curvature are the main reason for the information bottleneck between the graph and textual views (Refer to {\bf Theorem 2}). Therefore, we design a curvature-aware graph representation learning to enhance the textual and graph representation fusion. Our representation learning contains two steps: (1) the positive curvature-aware message transport and (2) the enhanced graph aggregation refining. To calculate the curvature for text and graph alignment, we first project textual and graph embeddings into the same curvature space. The above process for projected representations $\boldsymbol{\hat{X}}^{txt} \in \mathbb{R}^{N \times d}$ and $\boldsymbol{\hat{X}}^{gra} \in \mathbb{R}^{N \times d}$ can be written as
\begin{equation}
\begin{aligned}
\boldsymbol{\hat{X}}^{txt} = MLP(\boldsymbol{X}^{txt} | \boldsymbol{\theta}_t), \quad \boldsymbol{\hat{X}}^{gra} = MLP(\boldsymbol{X}^{gra} | \boldsymbol{\theta}_g).
\label{Projector_curvature}
\end{aligned}
\end{equation}
Here, $\boldsymbol{\theta}_t$ and $\boldsymbol{\theta}_g$ are the learnable parameters in the specific multilayer perceptron (MLP).

Because the negative curvature may hinder the information transport, our \methodbase considers aligning the textual and graph representations based on the positive curvatures. Specifically, in our first step of positive curvature-aware message transport, we extend the concept of curvature in Equation \eqref{ricii_curvature} to the scenarios for graph-driven LLMs. Since the probability measure $\boldsymbol{m}$ in curvature is used for neighbor information, we can use the one-hop graph propagation to aggregate the nodes of neighbors, where $\boldsymbol{m}$ can be calculated by $ \boldsymbol{m}_i = \sum_{k \in \mathcal{N}(i)} \boldsymbol{A}_{ik} \boldsymbol{\hat{X}}^{txt}_k$ and $ \boldsymbol{m}_j = \sum_{k \in \mathcal{N}(j)} \boldsymbol{A}_{jk} \boldsymbol{\hat{X}}^{gra}_k$ for node $i$ and $j$ under textual and graph views, respectively. Then, the Equation \eqref{ricii_curvature} can be reformulated:
\begin{equation}
\begin{aligned}
\boldsymbol{\kappa}_{ij} = 1 - \frac{W(\boldsymbol{m}_i, \boldsymbol{m}_j)}{d(\boldsymbol{\hat{X}}^{txt}_i, \boldsymbol{\hat{X}}^{gra}_j)}, 
\label{graph_ricii_curvature}
\end{aligned}
\end{equation}
where $W(\boldsymbol{m}_i, \boldsymbol{m}_j)$ is called the Wasserstein distance (or Earth Mover distance), which is often exploited in the field of optimal transport \cite{DBLP:conf/nips/Wu0ZWL24} or multi-modal alignment \cite{lin2023contrastive}. The Wasserstein distance can be obtained by solving the following optimal transportation plan:
\begin{equation}
\label{optimal_transport}
\begin{aligned}
\min_{M} & \sum_{i,j} d(\boldsymbol{m}_i, \boldsymbol{m}_j) M(\boldsymbol{m}_i, \boldsymbol{m}_j) \\
\text{s.t. } & \sum_j M(\boldsymbol{m}_i, \boldsymbol{m}_j) = 1/N, \forall i; \\
& \sum_i M(\boldsymbol{m}_i, \boldsymbol{m}_j) = 1/N, \forall j.
\end{aligned}
\end{equation}
Here, $M(\boldsymbol{x}_i, \boldsymbol{x}_j)$ is the optimal transportation plan from textual node $i$ to graph node $j$ and $d(\boldsymbol{m}_i, \boldsymbol{m}_j)$ is the geodesic distance. In the  {\bf Theorem 2}, we filter the negative curvatures and perform the inter-view message passing only when the positive curvature condition is satisfied($d(\boldsymbol{\hat{X}}^{txt}_i, \boldsymbol{\hat{X}}^{gra}_j) \textgreater W(\boldsymbol{m}_i, \boldsymbol{m}_j)$). Therefore, we design a novel transport mechanism while filtering the edge with negative curvatures as 
\begin{equation}
\begin{aligned}
\boldsymbol{T}_{ij}= ReLU(d(\boldsymbol{\hat{X}}^{txt}_i, \boldsymbol{\hat{X}}^{gra}_j) - W(\boldsymbol{m}_i, \boldsymbol{m}_j)),
\label{new_transport}
\end{aligned}
\end{equation}
where $\boldsymbol{T} \in \mathbb{R}^{N \times N}$ is the new transportation plan from textual nodes to graph nodes while filtering the negative curvatures and $ReLU(\cdot)$ is the activation function of Rectified Linear Unit (ReLU). Then, we obtain aligned fusion representations by $\boldsymbol{\hat{X}} = \lambda\boldsymbol{\hat{X}}^{gra} +  (1-\lambda)\boldsymbol{T}\boldsymbol{\hat{X}}^{txt}$.

Although the above fusion mechanism can effectively inject positive curvature information, the direct filtering of $edge(i,j)$ with negative curvature inevitably destroys the edge connection information between nodes $i$ and $j$. To alleviate the problem, we propose the enhanced graph aggregation refining. Specifically, after obtaining the fusion representations $\boldsymbol{\hat{X}} \in \mathbb{R}^{N \times d}$, we perform the graph aggregation for each node to refine the node representations by
\begin{equation}
\begin{aligned}
\boldsymbol{X}^{f} = GNN(
\boldsymbol{\hat{X}}, \boldsymbol{A} | \boldsymbol{\theta}),
\label{FM:GCN}
\end{aligned}
\end{equation}
where $\boldsymbol{\theta}$ is the trainable parameter in the graph neural network (GNN). The explanation of Equation \eqref{FM:GCN} is: For $edge(i,j)$ with negative curvature, the message between $i$ and $j$ can be seen as a two-stage updating, where the node $i$ can first transfer the information to the intermediate node $k$ with positive curvature, and then the infomation may be transferred to node $j$ through graph convolution ($i \xrightarrow{Equation \eqref{new_transport}} k \xrightarrow{Equation \eqref{FM:GCN}} j$). After the graph refining, we can obtain our final textual-graph fusion representation $\boldsymbol{X}^f \in \mathbb{R}^{N \times d}$, which is fed to the different heads for various downstream tasks. Here we take the node classification task as an example, where the head is an $MLP(\cdot)$ function. The above process can be written as
\begin{equation}
\begin{aligned}
\mathcal{L} = CrossEntropy(\boldsymbol{Y}, \boldsymbol{\hat{Y}}) \quad s.t. \quad \boldsymbol{\hat{Y}} = head(\boldsymbol{X}^f) = \sigma(
\boldsymbol{{X}}^{f}\boldsymbol{W} + \boldsymbol{b}).
\label{PLM_Pretrained}
\end{aligned}
\end{equation}
The prediction matrix $\boldsymbol{\hat{Y}} \in \mathbb{R}^{N \times C}$ is the predicted results by our \methodbase, where $C$ is the number of categories. $\boldsymbol{W} \in \mathbb{R}^{d \times C}$ and $\boldsymbol{b}$ are the learnable weight and bias in $head(\cdot)$ function.

\begin{table*}[!t]
\centering
\caption{Accuracy(\%) of all compared methods on semi-supervised node classification task, where the best and runner-up performance are highlighted in bold and underlined respectively.}
\label{res_cf}
    \begin{tabular}{c||cccccc}
    \toprule
    Method       & Instagram             & Cora                  & Citeseer              & Photo                 & WikiCS                & \textit{Average}        \\ \midrule
    MLP          & 64.67 (0.31)          & 56.47 (0.97)          & 70.40 (1.08)          & 62.47 (0.25)          & 73.63 (0.35)          & 65.53          \\ \midrule
    GCN          & 64.54 (0.30)          & 79.53 (1.36)          & 69.78 (0.27)          & 83.01 (0.28)          & 79.51 (0.40)          & 75.27          \\
    SAGE         & 64.76 (0.45)          & 81.50 (0.69)          & 71.22 (0.24)          & 83.11 (0.29)          & 81.85 (0.29)          & 76.49          \\
    CGCN         & 65.67 (0.08)          & 81.50 (0.22)          & 69.60 (1.07)          & 64.74 (0.27)          & 80.95 (0.13)          & 72.49          \\
    DSR-GNN      & 64.52 (0.25)          & 76.77 (0.36)          & 71.11 (0.55)          & OOM                     & 78.91 (0.44)          & -              \\ \midrule
    BERT         & 62.42 (3.03)          & 62.36 (2.29)          & 67.02 (1.75)          & 73.57 (0.48)          & 81.18 (1.00)          & 69.31          \\
    DeBERTa      & 64.90 (1.52)          & 36.46 (9.55)          & 45.83 (17.2)          & 74.42 (0.38)          & 79.56 (2.90)          & 60.23          \\
    SentenceBERT & 62.75 (0.90)          & 62.07 (3.92)          & 67.08 (3.05)          & 73.38 (0.54)          & 80.61 (1.35)          & 69.18          \\ \midrule
    GraphAdapter & 67.45 (0.51)    & 78.03 (1.46)          & 70.50 (0.90)          & OOM                     & OOM                     & -              \\
    TAPE         & -                     & {\underline {84.91 (0.90)}}    & -                     & -                     & -                     & -              \\
    WalkLM       & 66.40 (0.29)          & 74.45 (3.19)          & 69.55 (1.17)          & 76.19 (0.21)          & {\underline {84.14 (0.61)}}    & 74.15          \\
    ENGINE       & 66.93 (0.51)          & 81.75 (1.06)          & 72.32 (0.81)    & 84.64 (0.14)    & 84.04 (0.47)          & 77.94  \\ 
    STAG       & 66.98 (0.40)          & 82.17 (0.94)          & 71.09 (0.63)    & 83.26 (0.40)   & 84.08 (0.68)          & 77.52    \\
 MuseGraph       & \underline {{67.56 (0.34)}}          & 83.81 (1.14)          & {\underline {72.60 (0.72)}}    & {\underline {85.11 (0.25)}}    & 83.92 (0.53)          & {\underline {78.60}}    \\
    \midrule
    Ours         & \textbf{68.30 (0.01)} & \textbf{85.31 (0.09)} & \textbf{73.92 (0.29)} & \textbf{86.31 (0.07)} & \textbf{84.80 (0.05)} & \textbf{79.73} \\ \bottomrule
    \end{tabular}
\end{table*}

\section{Experiments}

In this section, we conduct extensive experiments to validate the effectiveness of the proposed \methodbase. The experiment is mainly divided into four parts and mainly focuses on answering the following four questions:

\textbf{RQ1:} How does the \methodbase perform in comparison
with other state-of-the-art models for graph-aware LLMs under the graph downstream tasks?

\textbf{RQ2: }How does each component proposed in the \methodbase  contribute to the performance improvement?

\textbf{RQ3: }How do the hyperparameters affect the prediction performance for the proposed \methodbase?

\textbf{RQ4: }How does the \methodbase improve the performance for modeling graph and textual information in graph-driven scenarios? 

\begin{table*}[!t]
\centering
\caption{Performance(\%) comparison of baselines in terms of Recall@10, and Precision@10 on Baby, Sports, MovieLen-1M, and MovieLen-10M datasets. We mark the global best results on each dataset under each metric in boldface and the second best is underlined.}
\label{res_lp}
\begin{tabular}{c|cc|cc|cc|cc}
\toprule
Dataset      & \multicolumn{2}{c}{Baby}                    & \multicolumn{2}{c}{Sports}                  & \multicolumn{2}{c}{MovieLen-1M}              & \multicolumn{2}{c}{MovieLen-10M}            \\ \midrule
Model        & R@10                 & P@10                 & R@10                 & P@10                 & R@10                 & P@10                  & R@10                 & P@10                 \\ \midrule
MF           & 3.76 (0.22)          & 0.29 (0.04)          & 4.81 (0.29)           & 0.38 (0.0)            & 5.63 (0.28)           & 9.11 (0.42)            & 9.34 (0.53)          & 9.13 (0.31)          \\
MLP          & 3.94 (0.24)          & 0.31 (0.03)          & 4.96 (0.25)           & 0.40 (0.08)           & 5.88 (0.28)           & 9.23 (0.46)            & 9.52 (0.55)           & 9.36 
 (0.30)          \\ \midrule
GCN          & 4.17 (0.19)          & 0.34 (0.04)          & 5.08 (0.27)           & 0.41 (0.07)           & 6.02 (0.29)           & 9.32 (0.49)            & 9.96 (0.44)           & 9.55 
 (0.31)          \\
GAT          & 4.31 (0.19)          & 0.38 (0.04)          & 5.16 (0.22)           & 0.44 (0.08)           & 6.23 (0.29)           & 9.58 (0.49)            & 10.21 (0.52)          & 10.07 
 (0.32)         \\
GIN          & 4.21 (0.21)          & 0.36 (0.03)          & 5.09 (0.28)           & 0.43 (0.08)           & 6.14 (0.25)           & 9.43 (0.40)            & 10.24 (0.39)          & 9.89 
 (0.29)          \\
SAGE         & 4.67 (0.15)          & 0.42 (0.03)          & 5.46 (0.23)           & 0.49 (0.06)           & 6.47 (0.28)           & 9.86 (0.43)            & 11.33 (0.41)          & 10.57 (0.30)         \\
GraphCL      & 4.86 (0.16)          & 0.45 (0.03)          & 5.54 (0.20)           & 0.52 (0.05)           & 6.81 (0.22)           & 9.92 (0.36)            & 12.11 (0.44)          & 10.63 
 (0.28)         \\ \midrule
Bert         & 4.03 (0.21)          & 0.32 (0.05)          & 5.09 (0.22)           & 0.45 (0.07)           & 6.10 (0.24)           & 9.37 (0.38)            & 10.06 (0.51)          & 9.42 (0.30)           \\
RoBERTa      & 4.22 (0.18)          & 0.36 (0.03)          & 5.24 (0.20)           & 0.51 (0.07)           & 6.22 (0.28)           & 9.42 (0.48)            & 10.37 (0.48)          & 9.92 (0.28)           \\ \midrule
WalkLM       & 4.56 (0.12)          & 0.39 (0.02)          & 5.56 (0.23)           & 0.55 (0.04)           & 7.25 (0.21)           & 10.01 (0.35)           & 12.66 (0.42)          & 10.68 (0.24)          \\
GraphAdapter & 4.95 (0.15)          & 0.48 (0.02)          & 5.94 (0.19)           & 0.59 (0.04)           & {\underline{ 7.65 (0.24)}}     & {\underline {10.18 (0.38)}}     & {OOM}    & {OOM}    \\
OpenGraph    & 5.01 (0.11)          & 0.49 (0.03)          & 5.82 (0.22)           & 0.53 (0.05)           & 7.03 (0.21)           & 9.98 (0.34)            & 12.88 (0.39)          & 10.74 (0.23)          \\
TAPE         & 5.03 (0.12)    & 0.52 (0.02)    & 6.14 (0.19)     &  0.61 (0.04)     & 7.44 (0.23)           & 10.12 (0.35)           & 13.68 (0.36)          & 10.81 (0.25)          \\ 
MuseGraph         & {\underline {5.09 (0.10)}}    & {\underline {0.54 (0.02)}}    & {\underline {6.25 (0.17)}}     & {\underline{ 0.66 (0.03)}}     & 7.58 (0.31)           & 10.15 (0.46)           & \underline{13.81 (0.43)}          & \underline{11.03 (0.31)}          \\ 
\midrule
Ours         & \textbf{5.33 (0.11)} & \textbf{0.59 (0.01)} & \textbf{6.62 (0.16)} & \textbf{0.74 (0.04)} & \textbf{8.60 (0.19)} & \textbf{11.05 (0.32)} & \textbf{14.95 (0.33)} & \textbf{12.18 (0.20)} \\ \bottomrule
\end{tabular}
\end{table*}

\subsection{Experimental Setup}
\textbf{Datasets.} We utilize a total of 11 real-world datasets, which are categorized according to three downstream graph learning tasks: 
(1) five datasets for node classification~\cite{graphadapter}, including the Cora, Citeseer, Instagram, Photo, and WikiCS datasets, to address common network relationships such as academic citations, social interactions, e-commerce, and encyclopedic knowledge; 
(2) four datasets for link prediction~\cite{opengraph}, including the Baby, Sports, ML-1M, and ML-10M datasets, which are customized for the e-commerce and entertainment recommendation domains containing user interaction attributes; 
(3) two datasets for graph question-answering (graphQA)~\cite{G-retriever}, namely the ExplaGraphs and WebQSP datasets, which focus on the knowledge reasoning domain and contain common-sense reasoning and large-scale multi-hop knowledge graph information requiring deep logical reasoning. 
These datasets have been widely adopted in previous literature~\cite{engine, wei2023multi, bm3, kong2024gofa, G-retriever}. For more details on dataset segmentation, feature dimensions, and relevant statistics, please refer to \textbf{Appendix C.1}.

\textbf{Implementation details.} For the classification task, following the existing settings \cite{tape}, we feed our learned fusion embeddings into a MLP head to obtain the final classification predictions. The widely-used \textit{Accuracy} metric is exploited to evaluate the classification performance. 
In link prediction task, we follow existing works \cite{wei2024multimodal}, \cite{opengraph} to conduct the full-rank test for each node, where the similarity of learned fusion embeddings can be calculated to predict the edge connection. To be specific, for each node, all nodes not connected to it in the training set are ranked by the model. The top-$K$ nodes are taken as positive predictions, and we calculate ${Recall@K}$ and ${Precision@K}$ scores with $K$ = $5$, $10$. 
%
For the GraphQA task, we follow the LLM head of existing work \cite{G-retriever}. Specifically,
the primary task of ExplaGraphs dataset is to assess whether
arguments are supportive or contradictory, using accuracy as the metric. For the WebQSP dataset, the $hit@1$ metric is used to assess the precision of the top returned answer. More implementation details are provided in {\bf Appendix C.2}.

\textbf{Baselines.} To verify the effectiveness of CureLLM, we compare it with traditional representaiton learning methods like MLP, MF, and the following representative and recent baselines, which can be mainly divided into three categories. \textbf{(1) GNN-based methods:} We select GCN \cite{gcn}, GAT \cite{GAT}, GIN \cite{GIN}, SAGE \cite{sage}, GraphCL \cite{GrapgCL}, CGCN \cite{CGCN}, and DSR-GNN \cite{DSR-GNN} as the compared methods.  \textbf{(2) Standard fine-tuned PLMs:} We select four widely used LMs: BERT(bert-base-uncased) \cite{bert}, DeBERTa (deberta-base) \cite{deberta}, SentenceBERT (bert-base-nli-mean-tokens) \cite{sentence-bert}, and RoBERTa (roberta-base) \cite{roberta}. \textbf{(3) LLM-enhanced methods:} We further compare the following state-of-the-art (SOTA) methods that combine LLMs and GNNs: WalkLM \cite{walklm}, TAPE \cite{tape}, OpenGraph \cite{opengraph}, ENGINE \cite{engine}, GraphAdapter \cite{graphadapter}, GraphToken \cite{graphtoken}, G-Retriever \cite{G-retriever}, STAG\cite{bo2025quantizing} and MuseGraph\cite{DBLP:journals/pami/TanLZWY26} and . Due to TAPE needs to obtain the interpretation of the text graph through GPT-3.5 and only the interpretation data on Cora is published. Therefore, we only report the classification results of this method on Cora. Besides, we also extend the baselines to our experiment setting of link prediction. Detailed descriptions for  the baselines can be found in {\bf Appendix C.3}.

\subsection{Overall Performance Comparison}
In order to evaluate the effectiveness of the proposed method, we compare the performance of our \methodbase framework to those of the baselines on different tasks which answers the \textbf{RQ1}.

\textbf{Node Classification.} We present the comparative analysis of node classification in Table \ref{res_cf}. 
As shown in Table 1, \methodbase consistently achieves the highest performance across all five node classification datasets (Instagram, Cora, Citeseer, Photo, and WikiCS). With an average accuracy of 79.73\%, the proposed method significantly outperforms all 14 baseline models. When compared to graph neural networks that rely on structural information, \methodbase yields substantial improvements; on the Cora dataset, for example, it elevates the accuracy from the 81.50\% achieved by SAGE to 85.31\%. This indicates that relying on topological structure leads to a representation bottleneck, making the incorporation of textual semantics crucial for enriching node features. Furthermore, against language models trained on node text attributes, our method secures considerable gains. On the WikiCS dataset, \methodbase surpasses BERT by an absolute margin of 3.62\%, proving that neglecting graph connectivity between nodes severely restricts classification accuracy. Finally, the model maintains a definitive advantage over text-graph fusion baselines. On the Citeseer dataset, it improves upon the second-best result of 72.60\% to reach 73.92\%, and on the Photo dataset, it advances from 85.11\% to 86.31\%. These results comprehensively demonstrate the efficacy of explicitly utilizing edge attributes to align the graph and text embeddings of identical nodes.

\textbf{Link Prediction.} We compare the link prediction results of our \methodbase to those of the baseline models. Table \ref{res_lp} shows the ${Recall@10}$ and ${Precision@10}$ on Baby, Sports, MovieLen-1M and MovieLen-10M datasets. In addition, we also conduct the experiments using the ${Recall@5}$ and ${Precision@5}$ metrics
and the results are provided in {\bf Appendix C.4}.
As observed in Table \ref{res_lp}, \methodbase outperforms all 14 baselines across all evaluation metrics on every dataset. When evaluated against graph neural networks relying on structural information, the proposed method yields significant improvements. For example, on the MovieLen-10M dataset, it elevates the Recall@10 of GraphCL from 12.11\% to 14.95\%. This demonstrates that effectively incorporating textual semantics significantly broadens the expressive capabilities of the model. Similarly, \methodbase secures substantial gains over pre-trained language models that rely on node textual attributes. On the MovieLen-1M dataset, for instance, it improves the Recall@10 of RoBERTa from 6.22\% to 8.60\%, proving that capturing higher-order topological relationships between nodes is a necessary prerequisite for accurate link prediction. Furthermore, when compared to the previously strongest baseline, MuseGraph, \methodbase demonstrates the most notable performance leap on the large-scale MovieLen-10M dataset, where Recall@10 increases from 13.81\% to 14.95\%, and Precision@10 rises from 11.03\% to 12.18\%. Unlike fusion models such as GraphAdapter, \methodbase successfully overcomes the memory constraints posed by large-scale graph data, completely avoiding out-of-memory (OOM) issues.
This demonstrates the effectiveness of the proposed method in link prediction tasks.

\begin{table}[!t]
    \centering
    \caption{Performance comparison across ExplaGraphs and WebQSP.}
    \label{res_qa}
    \begin{tabular}{ccc}
        \toprule
        Method & ExplaGraphs & WebQSP \\ \midrule
        Prompt tuning & 57.63 (2.43) & 48.34 (0.64) \\
        GraphToken & 85.08 (5.51) & 57.05 (0.74) \\
        G-Retriever & 85.16 (0.92) & 70.49 (1.21) \\ 
        MuseGraph & \underline{85.39 (0.83)} & \underline{70.61 (1.09)} \\
        \midrule
        Ours & \textbf{87.21 (0.85)} & \textbf{72.58 (0.51)} \\ \bottomrule
    \end{tabular}
\end{table}

\textbf{Graph Question Answering.} 
Table \ref{res_qa} presents a performance comparison across complex graph reasoning tasks. 
The proposed \methodbase outperforms traditional prompt tuning paradigms as well as the graph-aware baseline model GraphToken. Specifically, on the ExplaGraphs dataset, \methodbase achieves a state-of-the-art accuracy of 87.21\%, representing an absolute improvement of 1.82\% over the strongest baseline model, MuseGraph (85.39\%). Furthermore, on the more challenging WebQSP dataset, \methodbase improved the accuracy from the second-best result of 70.61\% to 72.58\%. 
This suggests that utilizing edge attributes to align the graph and textual embeddings is also effective for graphQA task.

Overall, across all datasets for the three tasks, our \methodbase outperforms all baseline models on all evaluation metrics. This answers RQ1, showing that mining edge attributes for aligning the graph and textual information is capable of learning a more effective representation.

\subsection{Ablation Study}
To answer the \textbf{RQ2} and provide the performance comparison of each component in \methodbase, we ablate each component by continuously adding different modules. 
We evaluate all components by removing each item from \methodbase in turn: (a) \textit{w/o} structural information: replacing GNN with MLP; (b) \textit{w/o} semantic information: removing textual embedding; (c) \textit{w/o} inference ability: replacing Llama2 with DeBERTa; (d) \textit{w/o} edge attribute: removing curvature-based alignment. The experiment results of node classification are presented in Table \ref{ablation_cf} and more experiment results can be found in {\bf Appendix C.5}. From Table \ref{ablation_cf}, we can observe that removing any components may lead to a decrease in performance.

\begin{table}[!t]
    \centering
    \setlength{\tabcolsep}{2pt}
    \caption{Ablation study of CureLLM on node classification where w/o denotes the removal of a specific component.}
    \label{ablation_cf}
    \resizebox{\columnwidth}{!}{
    \begin{tabular}{l|ccccc} \toprule
           & Instagram & Cora & Citeseer & Photo & WikiCS \\ \midrule
        \textit{w/o} structural information & 64.51 & 72.88 & 69.28 & 73.25 & 81.46 \\
        \textit{w/o} semantic information & 64.76 & 81.50 & \underline{71.22} & 83.11 & 81.85 \\
        \textit{w/o} inference ability & \underline{67.33} & 73.25 & 61.91 & 84.92 & 81.25 \\
        \textit{w/o} edge attribute & 66.84 & \underline{83.03} & 70.85 & \underline{85.67} & \underline{83.34} \\ \midrule
        CureLLM & \textbf{68.30} & \textbf{85.31} & \textbf{73.92} & \textbf{86.31} & \textbf{84.80} \\ \bottomrule
    \end{tabular}
    }
\end{table}



\begin{figure}[!t]
    \centering
\includegraphics[width=0.9\linewidth]{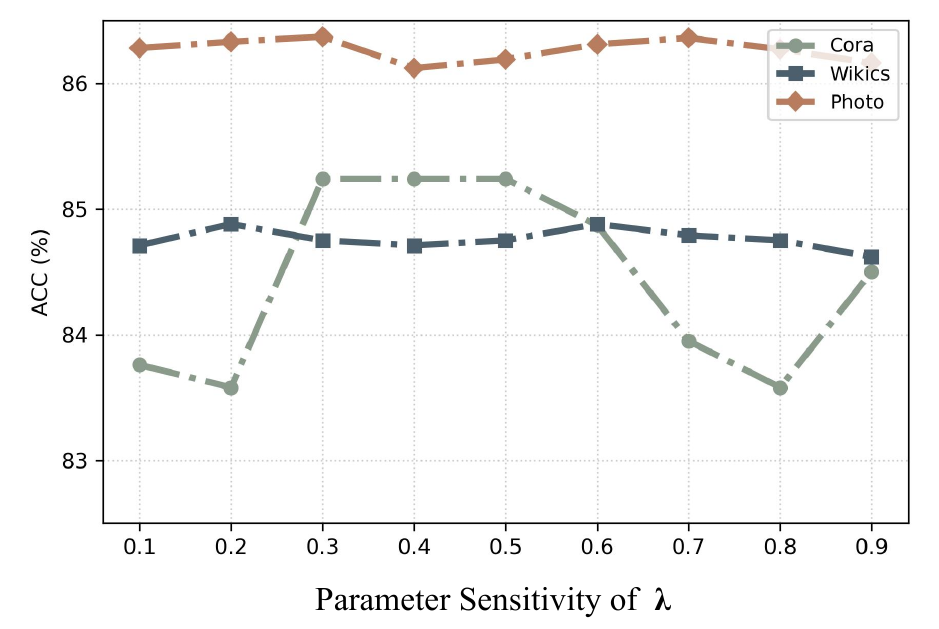}
    \caption{Parameter sensitivity analysis of $\lambda$ of \methodbase.}
    \label{fig:sens}
\end{figure}

\subsection{Parameter Sensitivity}
In this subsection, we will investigate how the parameter $\lambda$ impact \methodbase, which can answer \textbf{RQ3}. The parameter $\lambda$ is chosen from $0.1$ to $0.9$ with step size $0.1$. As shown in Figure \ref{fig:sens}, \methodbase attains optimal performance with $\lambda = 0.3$ or $\lambda = 0.5$ on most datasets. From the figure, we can observe that the proposed method is robust on photo dataset and wikics dataset. Combined with the results of the ablation experiments, it can be seen that the performance degrades when either the semantic information or the structural information is removed, indicating that the parameter is meaningful. The variations in the performance of \methodbase are evident on the Cora, Citeseer and Instagram. From the figure, we observe that whether $\lambda$ is too large or too small will make the performance sub-optimal. This suggests that focusing too much on the representation of a specific domain will prevent the model from learning the optimal representation.

\subsection{Case Study}
In this section of the case study, we aim to analyze and provide the explainable cases for different large language models to answer the question about \textbf{RQ4}.
To further validate the extensibility and architectural compatibility of the framework, we evaluated CureLLM’s performance across various combinations of large language models (LLMs) and graph neural networks (GNNs). For more details about results, please refer to \textbf{Appendix C.6}.


\vspace{-1mm}
\section{Conclusion}

 In this work, we alleviate the problem of partial alignment in the existing graph-aware large language model from the perspective of edge structure for the first time. We showed that current node-level alignment paradigms suffer from insufficient modeling of structural information, which leads to suboptimal information propagation across views. Through both theoretical analysis and empirical observations, we identified two key factors underlying this limitation: the neglect of edge-level structural information and the information compression induced by node-level aggregation. To address the challenges, we proposed a novel \methodbase framework whose goal is to exploit the signals of edge structure to enhance the existing alignment mechanism. Specifically, we design a training-free textual prompt mechanism to inject edge-aware structural information into pre-trained LLMs without introducing additional training cost. Furthermore, we develop a curvature-aware graph representation learning module, where curvature is leveraged as a geometric tool to characterize information propagation. Extensive experiments show the superiority of our proposed \methodbase.

\bibliographystyle{ACM-Reference-Format}
\bibliography{myref}

\clearpage
\appendix
\section{Appendix A for Proofs of Theorems and Related Work Discussion}

In this section, we will provide detailed proof of all our Theorems. {\bf Theorem 1.} is to illustrate the existing paradigm for graph LLMs using only node alignment while ignoring the structural information of edges can lead to suboptimal solutions. {\bf Theorem 2.} is to illustrate negatively curved edges are those
causing the graph bottleneck and thus leading to the partially indistinguishable phenomenon. Then, we will discuss the related work and give the prompt design of our \methodbase.

\subsection{Proofs for Theorem 1}

Following the definition of Generalization Error Upper Bound in \cite{DBLP:conf/icml/CaoXDZH24}, we can rewrite the formulation of generalization error as 

\begin{equation}
\begin{aligned}
GE(f)
=\mathbb{E}_{(x,y)\sim\mathcal{D}}\bigl[\ell(f(x),y)\bigr],
\label{definition_ge}
\end{aligned}
\end{equation}
where $\mathcal{D}$ is the selected dataset. By considering the property of the convex function as

\begin{equation}
\begin{aligned}
\ell(f(x), y) &= \ell\Bigl(\omega_n f_n(x_{(n)}) + \omega_e f_e(x_{(e)}), y\Bigr) \\
&\le \omega^m \ell(f^m(x^{(m)}), y) + \omega^m \ell(f^m(x^{(m)}), y)
\end{aligned}
\end{equation}
we can reformulate the Equation \ref{definition_ge} as 

\begin{equation}
GE(f) = \mathbb{E}_{(x,y)\sim\mathcal{D}} \bigl[ \ell(f(x),y) \bigr] \le \mathbb{E}_{(x,y)\sim\mathcal{D}} \bigl[ \omega_n\ell_n \bigr] + \mathbb{E}_{(x,y)\sim\mathcal{D}} \bigl[ \omega_e\ell_e \bigr].
\label{qiwan_bianhua}
\end{equation}
Here, $\omega_n$ and $\omega_e$ are the confidence weights for nodes and edges, respectively. The symbols of $\ell_n$ and $\ell_e$ are convex logistic loss functions for nodes and edges. Then, we make an equivalent transformation to the above Equation \ref{qiwan_bianhua} as

\begin{equation}
\begin{aligned}
GE(f)
&\le \mathbb{E}_{(x,y)\sim\mathcal{D}}\bigl[\omega_n\ell_n\bigr] + \mathbb{E}_{(x,y)\sim\mathcal{D}}\bigl[\omega_e\ell_e\bigr] \\
&= \frac{1}{2} \cdot \bigl(2 \cdot ( \mathbb{E}_{(x,y)\sim\mathcal{D}}\bigl[\omega_n\ell_n\bigr] + \mathbb{E}_{(x,y)\sim\mathcal{D}}\bigl[\omega_e\ell_e\bigr]) \bigr).
\label{qiwan_bianhua_equal}
\end{aligned}
\end{equation}
Since the weights of $\omega_n$ and $\omega_e$ satisfy the condition $(\omega_n + \omega_e = 1)$, we can rewrite Equation \ref{qiwan_bianhua_equal} as 

\begin{equation}
\begin{aligned}
GE(f)
&\le \frac{1}{2} \cdot \bigl(2 \cdot ( \mathbb{E}_{(x,y)\sim\mathcal{D}}\bigl[\omega_n\ell_n\bigr] + \mathbb{E}_{(x,y)\sim\mathcal{D}}\bigl[\omega_e\ell_e\bigr]) \bigr) \\
&= \frac{1}{2}\;\cdot \Bigl[ \Bigl(\mathbb{E}_{(x,y)\sim\mathcal{D}}\bigl[\omega_n\ell_n\bigr] 
+ \,\mathbb{E}_{(x,y)\sim\mathcal{D}}\bigl[(1-\! \omega_e)\,\ell_n\bigr]\Bigr)  \\
&+ \Bigl(\mathbb{E}_{(x,y)\sim\mathcal{D}}\bigl[\omega_e\ell_e\bigr] 
+ \,\mathbb{E}_{(x,y)\sim\mathcal{D}}\bigl[(1-\!\omega_n)\,\ell_e\bigr]\Bigr) \Bigr]
.
\label{qiwan_bianhua_equal_weight_1}
\end{aligned}
\end{equation}
Considering the definition of the covariance between the confidence weight $\omega$ and the loss $\ell$ by $Cov(\omega, \ell) = \mathbb{E}_{(x,y)\sim\mathcal{D}}\bigl[\omega\ell\bigr] - \mathbb{E}_{(x,y)\sim\mathcal{D}}\bigl[\omega\bigr] \cdot \mathbb{E}_{(x,y)\sim\mathcal{D}}\bigl[\ell\bigr] $, we can use the covariance to reformulate the Equation \ref{qiwan_bianhua_equal_weight_1} as 

\begin{equation}
\begin{aligned}
GE(f)
&\le \frac{1}{2} \Bigl[  \mathbb{E}_{(x,y)\sim\mathcal{D}}[\omega_n]
  \,\mathbb{E}_{(x,y)\sim\mathcal{D}}[\ell_n]
  +\mathrm{Cov}(\omega_n,\ell_n) \\
&+ \mathbb{E}_{(x,y)\sim\mathcal{D}}[\omega_e]
  \,\mathbb{E}_{(x,y)\sim\mathcal{D}}[\ell_e]
  +\mathrm{Cov}(\omega_e,\ell_e) \\
&- \Bigl(\mathbb{E}_{(x,y)\sim\mathcal{D}}[\omega_e]
  \,\mathbb{E}_{(x,y)\sim\mathcal{D}}[\ell_n]
  \\&+\mathrm{Cov}(\omega_e,\ell_n)
  -\mathbb{E}_{(x,y)\sim\mathcal{D}}[\ell_n]\Bigr) \\
&- \Bigl(\mathbb{E}_{(x,y)\sim\mathcal{D}}[\omega_n]
  \,\mathbb{E}_{(x,y)\sim\mathcal{D}}[\ell_e]
  \\&+\mathrm{Cov}(\omega_n,\ell_e)
  -\mathbb{E}_{(x,y)\sim\mathcal{D}}[\ell_e]\Bigr)
  \Bigr].
\label{qiwan_bianhua_equal_weight_reformulate}
\end{aligned}
\end{equation}
Then, we extend Equation \ref{qiwan_bianhua_equal_weight_reformulate} to

\begin{equation}
\begin{aligned}
GE(f) &\le \frac{1}{2} \biggl[ \Bigl( \mathbb{E}[\ell_n] \bigl(\mathbb{E}[\omega_n] + [1 - \mathbb{E}[\omega_e]]\bigr) \\
&\quad + \mathrm{Cov}(\omega_n, \ell_n) - \mathrm{Cov}(\omega_e, \ell_n) \Bigr) \\
&\quad + \Bigl( \mathbb{E}[\ell_e] \bigl(\mathbb{E}[\omega_e] + [1 - \mathbb{E}[\omega_n]]\bigr) \\
&\quad + \mathrm{Cov}(\omega_e, \ell_e) - \mathrm{Cov}(\omega_n, \ell_e) \Bigr) \biggr],
\label{qiwan_bianhua_equal_weight_extention}
\end{aligned}
\end{equation}
where we continue to simplify Equation \ref{qiwan_bianhua_equal_weight_extention} to obtain

\begin{equation}
\begin{aligned}
GE(f)
&\le \frac{1}{2} \Bigl[  \Bigl(
2\mathbb{E}_{(x,y)\sim\mathcal{D}}[\ell_n]
  \,\mathbb{E}_{(x,y)\sim\mathcal{D}}[\omega_n]
  \\&+\mathrm{Cov}(\omega_n,\ell_n)
  -\mathrm{Cov}(\omega_e,\ell_n) \Bigr) \\
&+  \Bigl(
2\mathbb{E}_{(x,y)\sim\mathcal{D}}[\ell_e]
  \,\mathbb{E}_{(x,y)\sim\mathcal{D}}[\omega_e]
  \\&+\mathrm{Cov}(\omega_e,\ell_e)
  -\mathrm{Cov}(\omega_n,\ell_e) \Bigr)
  \Bigr].
\label{qiwan_bianhua_equal_weight_simplfy}
\end{aligned}
\end{equation}
After the calculation of Equation \ref{qiwan_bianhua_equal_weight_simplfy}, we can have the final formulation of our Error Upper Bound for Graph Representation Learning in LLMs, shown as

\begin{equation}
\begin{aligned}
GE(f)
&\le  \mathbb{E}_{(x,y)\sim\mathcal{D}}[\ell_n]
  \,\mathbb{E}_{(x,y)\sim\mathcal{D}}[\omega_n]
  \\&+\frac{1}{2}\bigl[\mathrm{Cov}(\omega_n,\ell_n)
  -\mathrm{Cov}(\omega_n,\ell_e)\bigr] \\
&+ \mathbb{E}_{(x,y)\sim\mathcal{D}}[\ell_e]
  \,\mathbb{E}_{(x,y)\sim\mathcal{D}}[\omega_e]
  \\&+\frac{1}{2}\bigl[\mathrm{Cov}(\omega_e,\ell_e)
  -\mathrm{Cov}(\omega_e,\ell_n)\bigr].
\label{qiwan_bianhua_equal_weight_final}
\end{aligned}
\end{equation}
Then, we analyze the upper bound of Equation \ref{qiwan_bianhua_equal_weight_final}. Since the weights of $\omega_n$ and $\omega_e$ are less than 1, where it means $0 \le \omega_n \le 1$ and $0 \le \omega_e \le 1$, we can obtain the inequalities $\mathbb{E}_{(x,y)\sim\mathcal{D}}[\omega_n] \le 1$ and $\mathbb{E}_{(x,y)\sim\mathcal{D}}[\omega_e] \le 1$. Therefore, we can relax the upper bound of Equation \ref{qiwan_bianhua_equal_weight_final}, as shown as 

\begin{equation}
\begin{aligned}
GE(f)
&\le  \mathbb{E}_{(x,y)\sim\mathcal{D}}[\ell_n]
  +\frac{1}{2}\,\mathrm{Cov}(\omega_n,\ell_n)
  -\frac{1}{2}\mathrm{Cov}(\omega_n,\ell_e) \\
&+ \mathbb{E}_{(x,y)\sim\mathcal{D}}[\ell_e]
  +\frac{1}{2}\,\mathrm{Cov}(\omega_e,\ell_e)
  -\frac{1}{2}\mathrm{Cov}(\omega_e,\ell_n).
\label{qiwan_bianhua_equal_weight_upper_bound}
\end{aligned}
\end{equation}
Following the previous theory in Rademacher complexity \cite{trappenberg2018fundamentals}, the expectation of the above loss function has the following property: With a confidence level of $1-\Delta$ and $\Delta$ satisfies $0 < \Delta < 1 $, we get

\begin{equation}
\begin{aligned}
\mathbb{E}_{(x,y)\sim\mathcal{D}}[\ell_n]
& \leq  \mathcal{R}_N(\mathcal{H}) + \sqrt{ \frac{\ln(1/\Delta)}{2N} } 
+ \hat{{e}}(f).
\label{condition_upper_bound}
\end{aligned}
\end{equation}
Here, $\hat{{e}}(f)$ is the empirical error of the graph aggregation function $f(\cdot)$ and $\mathcal{H}$ is the hypothesis set. Besides, the Rademacher complexity is denoted by $\mathcal{R}_N(\mathcal{H})$. By combining the Equation \ref{qiwan_bianhua_equal_weight_upper_bound} and Equation \ref{condition_upper_bound}, the can obtain the final Error Upper Bound for Graph Representation Learning in LLMs with
\begin{equation}
\begin{aligned}
GE(f) & \leq  2\mathcal{R}_N(\mathcal{H}) \\&+ \sqrt{ \frac{2\ln(1/\Delta)}{N} } 
+ \hat{{e}}(f_n) + \hat{{e}}(f_e) \\ 
& + \frac{1}{2} \Bigg[ \underbrace{Cov(\omega_n, \ell_n)}_{\text{Node Covariance}} +
 \underbrace{ Cov(\omega_e, \ell_e)}_{\text{Edge Covariance}} \\&\underbrace{- Cov(\omega_n, \ell_e) - Cov(\omega_e, \ell_n)}_{{\text{Node-Edge Covariance}}}  \Bigg].
\end{aligned}
\end{equation}
Up to here, we have completed the proof of { \bf Theorem 1.}

\subsection{Proofs for Theorem 2}

In the { \bf Theorem 2}, our goal is to prove the following conclusion:

{\bf Theorem 2.} (Negative Curvature in Graph Representation Learning with LLMs) \textit{Consider message passing and curvature. Let node $i$ under the graph view and node $j$ under the textual view with $d_i \leq d_j$. Assume that the derivatives of functions under graph-graph and text-text message passing are bounded with $|\nabla \phi_l| \leq \alpha$ \textit{and} $|\nabla \psi_l| \leq \beta$, and the message passing functions between graph-text nodes are bounded with $|\nabla \phi_l^{\prime}| \leq \alpha^{\prime}$ \textit{and} $|\nabla \psi_l^{\prime}| \leq \beta^{\prime}$, \textit{for each} $0 \leq l \leq L-1$, \textit{with} the depth of message passing $L \geq 2$. \textit{There exists $\delta$ and $\gamma_{\text{max}}^{-1}$ such that } $0 < \delta < \left(\max\{d_i, d_j\}\right)^{-\frac{1}{2}},\ \delta < \gamma_{\text{max}}^{-1}$ \textit{and } $\boldsymbol{\kappa}_{ij} \leq -2 + \delta$.}
\textit{Then given $Q_j \subseteq S_2(i)$ satisfying $|Q_j| > \delta^{-1}$ and for $0 \leq l_0 \leq L - 2$, we have}
\begin{equation}
\begin{aligned}
\frac{1}{|Q_j|} \sum_{k \in Q_j} \left| \frac{\partial \boldsymbol{h}_k^{(l_0 + 2)}}{\partial \boldsymbol{h}_i^{(l_0)}} \right| < (\alpha \beta) ( \alpha^{\prime} \beta^{\prime}) \delta^{\frac{1}{4}}.
\label{negative_ricii_curvature}
\end{aligned}
\end{equation}
To prove the above conclusion, we need to first define the geometric sign of curvature. We let the degree of node $i$ and $j$ in curvature $edge(i,j)$ as $d_i := d$ and $d_j := d+s$ with $s > 0$. Then, the symbol $tri(i,j)$ is set of the triangles based at nodes $i$ and $j$, where the set $tri(i,j)$ is defined as $tri(i,j) := S_{1}(i)\,\cap\,S_{1}(j)$. The symbol $squ(i,j)$ means the set of 4-cycle squares based at the edge $edge(i,j)$, where the set $squ^i(i,j)$ is defined as $squ^i(i,j) := \{\,k\in S_1(i)\setminus S_1(j),\;k\neq j:\;\exists\,w\in\bigl(S_1(k)\cap S_1(j)\bigr)\setminus S_1(i) \}$. Here, both $tri(i,j)$ and $squ(i,j)$ are the common definitions in the curvature from the geometric perspective, e.g. \cite{ollivier2007ricci, ollivier2009ricci}. Then, we introduce the { \bf Lemma 1} to help the proof for {\bf Theorem 2}:

{\bf Lemma 1.}\textit{
Assume a graph message passing mechanism with the mechanism of  $\boldsymbol{h}_i^{(l+1)} = \phi_l ( \boldsymbol{h}_i^{(l)}, \sum_{j=1}^{n} \boldsymbol{A}_{ij} \psi_l ( \boldsymbol{h}_i^{(l)}, \boldsymbol{h}_j^{(l)} ) ).$ Let \(i,s\in V\) with \(s\in S_{r+1}(i)\). If $\bigl\lvert\nabla\phi_\ell\bigr\rvert\le\alpha
\quad\text{and}\quad
\bigl\lvert\nabla\psi_\ell\bigr\rvert\le\beta
\quad\text{for }0\le\ell\le r$, then we can have:} 
\begin{equation}
\begin{aligned}
\left\lvert
\frac{\partial \boldsymbol{h}_i^{(r+1)}}{\partial \boldsymbol{x}_s}
\right\rvert
\le(\alpha\beta)^{r} ( \alpha^{\prime} \beta^{\prime}) \bigl(\boldsymbol{\hat A}^{\,r+1}\bigr)_{is}\,.
\label{Llama_1}
\end{aligned}
\end{equation}
To prove { \bf Lemma 1}, we first calculate the derivative expansion of Equation \ref{Llama_1}. According to the chain rule of multivariate function:
\begin{equation}
\begin{aligned}
\frac{\partial \boldsymbol{h}_i^{(r+1)}}{\partial \boldsymbol{x}_s}
=
\partial_1\phi_r\bigl(\boldsymbol{u}_i,\boldsymbol{v}_i\bigr)\,\frac{\partial \boldsymbol{u}_i}{\partial \boldsymbol{x}_s}
\;+\;
\partial_2\phi_r\bigl(\boldsymbol{u}_i,\boldsymbol{v}_i\bigr)\,\frac{\partial \boldsymbol{v}_i}{\partial \boldsymbol{x}_s},
\label{Llama_1_multivariate}
\end{aligned}
\end{equation}
the derivative expansion of Equation \ref{Llama_1} is shown as
%
\begin{equation}
\begin{aligned}
&\frac{\partial \boldsymbol{h}_i^{(r+1)}}{\partial \boldsymbol{x}_s} = \partial_1\phi_r\bigl(\dots\bigr)\,\partial_{\boldsymbol{x}_s}\boldsymbol{h}_i^{(r)} \\
&+\partial_2\phi_r\bigl(\dots\bigr) \sum_{j_r=1}^n \hat a_{i j_r} \Bigl[ \partial_1\psi_r\bigl(\boldsymbol{h}_i^{(r)},\boldsymbol{h}_{j_r}^{(r)}\bigr)\,\partial_{\boldsymbol{x}_s}h_i^{(r)} \\
&\quad + \partial_2\psi_r\bigl(\boldsymbol{h}_i^{(r)},\boldsymbol{h}_{j_r}^{(r)}\bigr)\,\partial_{\boldsymbol{x}_s}\boldsymbol{h}_{j_r}^{(r)} \Bigr].
\label{Llama_1_multivariate_expansion}
\end{aligned}
\end{equation}
We can iterate the above computation and the right-hand side of Equation \ref{Llama_1_multivariate_expansion} can be expanded as
\begin{equation}
\begin{aligned}
&\frac{\partial \boldsymbol{h}_i^{(r+1)}}{\partial \boldsymbol{x}_s} = \sum_{j_r,\dots,j_0} \sum_{k_r\in\{i,j_r\}} \dots \sum_{k_1\in\{i,j_r,\dots,j_1\}} \\
&\quad \hat a_{i j_r} \hat a_{k_r j_{r-1}} \dots \hat a_{k_1 j_0} \, Z_{i j_r k_r j_{r-1} \dots k_1 j_0}(\boldsymbol{X}) \, \frac{\partial \boldsymbol{h}^{(0)}_{j_0}}{\partial \boldsymbol{x}_s}
\label{Llama_1_multivariate_expansion_iteration}
\end{aligned}
\end{equation}
where $Z_{\,i j_r k_r j_{r-1}\dots k_1 j_0}(\cdot)$ means the functions of input features obtained as products of $r+1$ partial derivatives for the message
function $\psi_l$ and the update function $\phi_l$. Since $\boldsymbol{H}^{(0)} = \boldsymbol{X}$ and we have 
\begin{equation}
\begin{aligned}
\frac{\partial \boldsymbol{h}^{(0)}_{j_0}}{\partial \boldsymbol{x}_s} = \delta_{j_0s},
\label{Llama_1_multivariate_expansion_iteration_we_have}
\end{aligned}
\end{equation}
where we can we can derive the following formulation:
%
\begin{equation}
\begin{aligned}
&\frac{\partial \boldsymbol{h}_i^{(r+1)}}{\partial \boldsymbol{x}_s} = \sum_{j_r,\dots,j_1} \sum_{k_r\in\{i,j_r\}} \dots \sum_{k_1\in\{i,j_r,\dots,j_1\}} \\
&\quad \hat a_{i j_r} \hat a_{k_r j_{r-1}} \dots \hat a_{k_1 s} \, Z_{i j_r k_r j_{r-1} \dots k_1 s}(\boldsymbol{X})
\label{Llama_1_multivariate_expansion_iteration_we_have_derive}
\end{aligned}
\end{equation}
Here, the only non-vanishing terms in the sum above are the minimal walks from $i$ to $s$ with $d_G(i,s) = r+1$. Besides, $Z_{\,i j_r k_r j_{r-1}\dots k_1 s}(\cdot)$ is a product of $r+1$-partial derivatives of the aggregation and update maps and by assumption their gradients, which are bounded by $\alpha$ and $\beta$.
Note that, compared with the common graph message passing mechanism, our \methodbase is tailored for the message propagation between the graph and textual nodes.  If we define the bridge node between graph and textual views as $t$, due to limited assumptions that he message passing functions between graph-text nodes are bounded with $|\nabla \phi_l^{\prime}| \leq \alpha^{\prime}$ \textit{and} $|\nabla \psi_l^{\prime}| \leq \beta^{\prime}$, we can have the final formulation as
\begin{equation}
\begin{aligned}
\left\lvert \frac{\partial \boldsymbol{h}_i^{(r+1)}}{\partial \boldsymbol{x}_s} \right\rvert
\;&\le\;
(\alpha\beta)^{\,r} ( \alpha^{\prime} \beta^{\prime})
\sum_{j_r,\dots,j_1}
\hat a_{i j_r}\,\hat a_{j_r j_{r-1}} \cdots \hat a_{j_{r-t} j_{r-t-1}}\,\cdots\,\hat a_{j_1 s}
\;
\\
&=\;
(\alpha\beta)^{r} ( \alpha^{\prime} \beta^{\prime}) \bigl(\boldsymbol{\hat A}^{\,r+1}\bigr)_{is}\,.
\label{Llama_1_multivariate_expansion_iteration_Final_formularion}
\end{aligned}
\end{equation}
Up to here, we have completed the proof for {\bf Lemma 1}.

Here, we begin the proof of Equation \ref{negative_ricii_curvature} in {\bf Theorem 2}. From the assumption of Equation \ref{negative_ricii_curvature}, we can derive 
$\boldsymbol{\kappa}_{ij} \leq -2 + \delta$ and $\delta^2(d+s) \le 1$ when the condition meets
\begin{equation}
\begin{aligned}
&4 + 2\frac{s}{d} + 3|tri(i,j)| + \frac{s}{d}|tri(i,j)| \\
&\quad + \gamma_{\max}^{-1} \bigl( |squ^i(i,j)| + |squ^j(i,j)| \bigr) \le \delta(d + s).
\label{negative_ricii_assumption_conditions}
\end{aligned}
\end{equation}
where $\gamma_{\max}^{-1}(i,j)$ means the maximal number of 4-cycles based at nodes $i$ and $j$ traversing a common node. Then, we can drive the following formulation:
\begin{equation}
\begin{aligned}
\delta\,\bigl|tri(i,j)\bigr|\Bigl(3 + \frac{s}{d}\Bigr)
\;\le\;\delta^{2}\,(d + s),
\label{negative_ricii_assumption_conditions_derive}
\end{aligned}
\end{equation}
with the constraint formulation shown as
\begin{equation}
\begin{aligned}
\delta\,\bigl|tri(i,j)\bigr| \le 1.
\label{negative_ricii_assumption_conditions_derive_constrain}
\end{aligned}
\end{equation}
Following the above conditions, we let $Q_j$ denote again the complement $S_{1}(j)\setminus\bigl(S_{1}(i)\,\cup\,squ^j(i,j)\,\cup\,\{i\}\bigr)$, where we also set $l_0 = 0$ and $\boldsymbol{h}_i^{(0)} = \boldsymbol{x}_i$ with $k \in Q_j$ and $k \in S_2(i)$. Then, we can rewrite the Equation \ref{Llama_1_multivariate_expansion_iteration_Final_formularion} in {\bf Lemma 1} as
\begin{equation}
\begin{aligned}
\left\lvert \frac{\partial \boldsymbol{h}_k^{(2)}}{\partial \boldsymbol{x}_i} \right\rvert
\;&\le\;
(\alpha\beta) ( \alpha^{\prime} \beta^{\prime}) \bigl(\boldsymbol{\hat A}\bigr)^2_{ik}\,,
\label{Llama_1_rewrite}
\end{aligned}
\end{equation}
where the definition of the normalized adjacency matrix $\boldsymbol{\hat A}$ is
\begin{equation}
\begin{aligned}
\bigl(\boldsymbol{\hat A}\bigr)^2_{ik}
= \frac{1}{\sqrt{(d_k + 1)\,(d_i + 1)}}
  \sum_{w \in S_1(k)\,\cap\,S_1(i)}
  \frac{1}{d_w + 1}\,.
\label{normalized_adjacency_matrix}
\end{aligned}
\end{equation}
If we set $\widehat Q_j \;=\;\bigl\{\,k \in Q_j : \sigma_{ik} > 1\bigr\}\,$, we can rewrite Equation \ref{normalized_adjacency_matrix} as
%
\begin{equation}
\begin{aligned}
&\sum_{k\in Q_j}(\boldsymbol{\hat A})^2_{ik} = \sum_{k\in Q_j} \frac{1}{\sqrt{(d_k+1)(d_i+1)}} \sum_{w\in S_1(k)\cap S_1(i)} \frac{1}{d_w+1} \\
&= \frac{1}{\sqrt{d_i+1}} \Biggl( \underbrace{\sum_{k\in Q_j} \frac{1}{\sqrt{d_k+1}}\frac{1}{d_j+1}}_{\text{First term}} \\
&\quad + \underbrace{\sum_{k\in \widehat Q_j} \frac{1}{\sqrt{d_k+1}} \sum_{w\in S_1(k)\cap S_1(i)\cap S_1(j)} \frac{1}{d_w+1}}_{\text{Second term}} \Biggr).
\label{normalized_adjacency_matrix_rewrite}
\end{aligned}
\end{equation}
Now, we first analyze the bound of the first term, from which we can conclude that:
\begin{equation}
\begin{aligned}
&\frac{1}{\sqrt{d_i+1}}
\sum_{k\in Q_j}
\frac{1}{\sqrt{d_k+1}}\;\frac{1}{d_j+1}
\;\le\;\\&
\frac{1}{\sqrt{d_i+1}}\;\lvert Q_j\rvert\;\frac{1}{d_j+1}
\;\le\;\\&
\frac{1}{\sqrt{d_i+1}}
\;\le\;
1.
\label{First_Term_Conclusion}
\end{aligned}
\end{equation}
Through the Equation \ref{First_Term_Conclusion}, we can know that the first term has a constant upper bound, and we analyze the bound of the second term. To simplify the derived mathematical symbols, we define 
\begin{equation}
\begin{aligned}
\Omega &:= \biggl\{w \in tri(i,j) : d_{w} < \frac{1}{C} \frac{\lvert \widehat Q_{j}\rvert}{\lvert tri(i,j)\rvert} + \frac{2}{C}\biggr\} \\
\text{and } V_k &:= S_{1}(k) \cap S_{1}(i) \cap S_{1}(j).
\label{derived_mathematical_symbols}
\end{aligned}
\end{equation}
Then, the second term can be split as
\begin{equation}
\begin{aligned}
\sum_{k\in \widehat Q_j}
\frac{1}{\sqrt{d_k+1}}\,
\Biggl(
  \sum_{w\in V_k\cap \Omega}\frac{1}{d_w+1}
  \;+\;
  \sum_{w\in V_k\setminus \Omega}\frac{1}{d_w+1}
\Biggr).
\label{Second_term_split}
\end{aligned}
\end{equation}
For the left-hand side of Equation \ref{Second_term_split} with $w\in V_k\cap \Omega$, we obtain the following inequality:
\begin{equation}
\begin{aligned}
\sum_{k\in \widehat Q_j}
\frac{1}{\sqrt{d_k+1}}
\Bigl(\sum_{w\in V_k\cap \Omega}\frac{1}{d_w+1}\Bigr)
\;\le\;
\sum_{k\in \widehat Q_j}
\frac{1}{\sqrt{d_k+1}}\;\frac{\lvert V_k\cap \Omega\rvert}{4}\,,
\label{Second_term_split_left_hand}
\end{aligned}
\end{equation}
and the observation
\begin{equation}
\begin{aligned}
\sum_{k\in \widehat Q_j}
\frac{\lvert V_k \cap \Omega\rvert}{\sqrt{d_k + 1}}
\;\le\;
\sum_{k\in \widehat Q_j}\lvert V_k \cap \Omega\rvert
\;=\;\\
\bigl\lvert\{(k,w)\in E : k\in \widehat Q_j,\;w\in V_k\cap \Omega\}\bigr\rvert
\;\le\;
\bigl(\max_{w\in\Omega}d_w\bigr)\,\lvert \Omega\rvert.
\label{Second_term_split_left_hand_observation}
\end{aligned}
\end{equation}
Since $d_w$ meets $d_w \;\le\; \frac{1}{C}\,\frac{\lvert \widehat Q_j\rvert}{\lvert tri(i,j)\rvert}
       \;+\;\frac{2}{C}$, we can conclude the left-hand side of Equation \ref{Second_term_split} can be bounded by
\begin{equation}
\begin{aligned}
\sum_{k\in \widehat Q_j}
\frac{1}{\sqrt{d_k+1}}
\sum_{w\in V_k\cap \Omega}
\frac{1}{d_w+1}
\;\le\;
\sum_{k\in \widehat Q_j}
\frac{1}{\sqrt{d_k+1}}
\;\frac{\lvert V_k\cap \Omega\rvert}{4}
\;
\\
\le\;
\Bigl(\frac{1}{C}\,\frac{\lvert \widehat Q_j\rvert}{\lvert tri(i,j)\rvert}
      +\frac{2}{C}\Bigr)\,\frac{\lvert \Omega\rvert}{4}
\;\le\;
\Bigl(\frac{1}{C}\,\frac{\lvert \widehat Q_j\rvert}{\lvert tri(i,j)\rvert}
      +\frac{2}{C}\Bigr)\,\frac{\lvert tri(i,j)\rvert}{4}.
\label{Second_term_split_left_hand_observation_bound}
\end{aligned}
\end{equation}
Then, we analyze the right-hand side of Equation \ref{Second_term_split} with $w\in V_k\setminus \Omega$. We can obtain the formulation
\begin{equation}
\begin{aligned}
\sum_{k\in \widehat Q_j}
\frac{1}{\sqrt{d_k+1}}
\sum_{w\in V_k\setminus \Omega}
\frac{1}{d_w+1}
\;\le\;
\sum_{k\in \widehat Q_j}
\frac{1}{\sqrt{d_k+1}}
\;\frac{C\,\lvert tri(i,j)\rvert}{\lvert \widehat Q_j\rvert}\,
\lvert V_k\setminus \Omega\rvert,
\label{Second_term_split_right_hand}
\end{aligned}
\end{equation}
where we can use the property $d_w^{-1} \;\le\; C\,\frac{\lvert tri(i,j) \rvert}{\lvert\widehat Q_j\rvert}$. Since the following inequalities
\begin{equation}
\begin{aligned}
\frac{\lvert V_k\setminus \Omega\rvert}{\sqrt{d_k+1}}
\;\le\;
\frac{\lvert V_k\rvert}{\sqrt{\lvert S_1(k)\rvert}}
\;\le\;
\frac{\lvert V_k\rvert}{\sqrt{\lvert V_k\rvert}}
\;\le\;
\sqrt{\lvert V_k\rvert}
\;\le\;
\sqrt{\lvert tri(i,j)\rvert}\,,
\label{Second_term_split_right_hand_inequality}
\end{aligned}
\end{equation}
the right-hand side of the Equation \ref{Second_term_split} can be bounded by
\begin{equation}
\begin{aligned}
\sum_{k\in \widehat Q_j}
\frac{1}{\sqrt{d_k+1}}
\;\frac{C\,\lvert tri(i,j)\rvert}{\lvert\widehat Q_j\rvert}
\;\bigl\lvert V_k\setminus\Omega\bigr\rvert
\;\le\;\\
\frac{C\,\lvert tri(i,j)\rvert}{\lvert\widehat Q_j\rvert}
\;\sqrt{\lvert tri(i,j)\rvert\,\lvert\widehat Q_j\rvert}
\;=\;
C\,\lvert tri(i,j)\rvert^{\tfrac{3}{2}}.
\label{Second_term_split_right_hand_bound}
\end{aligned}
\end{equation}
we have calculated the upper bounds of all intermediate terms of Equation \ref{normalized_adjacency_matrix_rewrite}. Specifically, the Equation \ref{normalized_adjacency_matrix_rewrite} can be termed as
\begin{equation}
\begin{aligned}
\frac{1}{\lvert Q_j\rvert}
\sum_{k\in Q_j}(\boldsymbol{\hat A})^2_{ik}
\;&\le\;
\frac{1}{\lvert Q_j\rvert}
\;+\;
\frac{1}{\lvert Q_j\rvert}
\Biggl(
  \sum_{k\in\widehat Q_j}
    \frac{1}{\sqrt{d_k+1}}
    \sum_{w\in V_k}\frac{1}{d_w+1}
\Biggr)
\;
\\
&\le\;
\delta
\;+\;
\frac{1}{\lvert Q_j\rvert}
\Biggl(
  \sum_{k\in\widehat Q_j}
    \frac{1}{\sqrt{d_k+1}}
    \sum_{w\in V_k}\frac{1}{d_w+1}
\Biggr),
\label{adj_matrix_bound}
\end{aligned}
\end{equation}

\begin{table*}[!h]
  \centering
  \caption{Prompts used in this work to query the \methodbase.}
  \label{tab:prompts}
  \small 
  \begin{tabular}{@{} l p{0.82\textwidth} @{}}
    \toprule
    \textbf{Dataset} & \textbf{Prompt} \\
    \midrule
    Instagram & 
    \textbf{Caption:} {\color{blue}<caption text>} | \textbf{Comment:} {\color{blue}<comment text>} | \textbf{Usertag:} {\color{blue}<tag text>} \par
    \textbf{Category:} {\color{blue}<category text>} \par
    \textbf{Question:} Based on the profile of the following Instagram account, please determine which category it belongs to and explain the reasons within 256 words. \textbf{Answer:} \\ \midrule
    
    Cora & 
    \textbf{Abstract:} {\color{blue}<abstract text>} | \textbf{Title:} {\color{blue}<title text>} | \textbf{Topic:} {\color{blue}<topic text>} \par
    \textbf{Question:} The following paper belongs to what category and explain the reasons within 256 words. \textbf{Answer:} \\ \midrule
    
    Citeseer & 
    \textbf{Abstract:} {\color{blue}<abstract text>} | \textbf{Title:} {\color{blue}<title text>} \par
    \textbf{Question:} The following paper belongs to what category and explain the reasons within 256 words. \textbf{Answer:} \\ \midrule
    
    Photo & 
    \textbf{Consumer:} {\color{blue}<consumer text>} | \textbf{Review:} {\color{blue}<review text>} \par
    \textbf{Question:} The following electronics products belong to what category and explain the reasons within 256 words. \textbf{Answer:} \\ \midrule
    
    WikiCS & 
    \textbf{Abstract:} {\color{blue}<abstract text>} | \textbf{Title:} {\color{blue}<title text>} \par
    \textbf{Question:} The following contents of the Wikipedia article belong to what category and explain the reasons within 256 words. \textbf{Answer:} \\\midrule

    Baby & 
    \textbf{Node ID:} {\color{blue}<id text>} | \textbf{Title:} {\color{blue}<title text>} | \textbf{Category:} {\color{blue}<category text>} | \textbf{Description:} {\color{blue}<description text>} | \textbf{Brand:} {\color{blue}<brand text>} \par
    \textbf{Question:} The following item may be bought by what kind of users and explain the reasons within 256 words. \textbf{Answer:} \\ \midrule
    
    Sports & 
    \textbf{Node ID:} {\color{blue}<id text>} | \textbf{Title:} {\color{blue}<title text>} | \textbf{Category:} {\color{blue}<category text>} | \textbf{Description:} {\color{blue}<description text>} | \textbf{Brand:} {\color{blue}<brand text>} \par
    \textbf{Question:} The following item may be bought by what kind of users and explain the reasons within 256 words. \textbf{Answer:} \\ \midrule
    
    ML-1M & 
    \textbf{MovieID:} {\color{blue}<movie id text>} | \textbf{Title:} {\color{blue}<title text>} | \textbf{Genre:} {\color{blue}<genre text>} \par
    \textbf{Question:} Based on the movie details, the following movie may be bought by what kind of users and explain the reasons within 256 words. \textbf{Answer:} \\ \midrule
    
    ML-10M & 
    \textbf{MovieID:} {\color{blue}<movie id text>} | \textbf{Title:} {\color{blue}<title text>} | \textbf{Genre:} {\color{blue}<genre text>} \par
    \textbf{Question:} Based on the movie details, the following movie may be bought by what kind of users and explain the reasons within 256 words. \textbf{Answer:} \\ \midrule
    
    ExplaGraphs & 
    \textbf{Node Attribute:} {\color{blue}<node attribute text>} | \textbf{Edge Attribute:} {\color{blue}<edge attribute text>} | \textbf{Argument:} {\color{blue}<argument text>} \par
    \textbf{Question:} Do argument 1 and argument 2 support or counter each other? Answer in one word in the form of support or counter, and explain the reasons within 256 words. \textbf{Answer:} \\ \midrule
    
    WebQSP & 
    \textbf{Node Attribute:} {\color{blue}<node attribute text>} | \textbf{Edge Attribute:} {\color{blue}<edge attribute text>} | \textbf{Description:} {\color{blue}<description text>} \par
    \textbf{Question:} Please answer the given question based on the knowledge graph, and explain the reasons within 256 words. \textbf{Answer:} \\
    \bottomrule
  \end{tabular}
\end{table*}

where we use the property that the $\lvert Q_j\rvert$ can be bounded by $\delta$. By considering the Equation \ref{Second_term_split_left_hand_observation_bound} and Equation \ref{Second_term_split_right_hand_bound} to estimate the second term in Equation \ref{normalized_adjacency_matrix_rewrite}, we can have
\begin{equation}
\begin{aligned}
&\frac{1}{\lvert Q_{j}\rvert}
\Bigl(
  \sum_{k\in \widehat Q_{j}}
    \frac{1}{\sqrt{d_{k}+1}}
    \sum_{w\in V_{k}}
      \frac{1}{d_{w}+1}
\Bigr)
\;\\
&\le\;
\frac{1}{\lvert Q_{j}\rvert}
\Bigl(
  \Bigl(\frac{1}{C}\,\frac{\lvert \widehat Q_{j}\rvert}{\lvert tri(i,j)\rvert}
       +\frac{2}{C}\Bigr)
  \frac{\lvert tri(i,j)\rvert}{4}
\Bigr)
+
\frac{1}{\lvert Q_{j}\rvert}
\bigl(C\,\lvert tri(i,j)\rvert^{\tfrac32}\bigr)
\;
\\
&\le\;
\frac{1}{4}
\Bigl(\frac{1}{C}
      +\frac{2\,\lvert tri(i,j)\rvert}{C\,\lvert Q_{j}\rvert}\Bigr)
+\frac{C\,\lvert tri(i,j)\rvert}{\lvert Q_{j}\rvert}
\sqrt{\lvert tri(i,j)\rvert}\,.
\label{adj_matrix_bound_second_term_est}
\end{aligned}
\end{equation}
Then, we continue to use the the property that the $\lvert Q_j\rvert$ can be bounded by $\delta$, the Equation \ref{adj_matrix_bound_second_term_est} can be termed as
\begin{equation}
\begin{aligned}
\frac{1}{4}\Bigl(\frac{1}{C}+\frac{2\,\lvert tri(i,j)\rvert}{C\,\lvert Q_{j}\rvert}\Bigr)
+\frac{C\,\lvert tri(i,j)\rvert}{\lvert Q_{j}\rvert}\sqrt{\lvert tri(i,j)\rvert}
\;\le\;\\
\frac{1}{4}\Bigl(\frac{1}{C}+\frac{2\,\delta}{C}\Bigr)
+ C\,\delta\,\sqrt{\lvert tri(i,j)\rvert}\,.
\label{adj_matrix_bound_second_term_est_use_propety}
\end{aligned}
\end{equation}
Now, we let $C = \delta^{-\frac{1}{4}}$ and perform constant multiple reduction for Equation \ref{adj_matrix_bound_second_term_est_use_propety}, which does not affect the final conclusion. Then, we can rewrite the Equation \ref{adj_matrix_bound_second_term_est_use_propety} as
\begin{equation}
\begin{aligned}
&\frac{1}{4}\Bigl(\frac{1}{C}+\frac{2\delta}{C}\Bigr)
+ C\,\delta\,\sqrt{\lvert tri(i,j)\rvert}\\
&\;\le\;
\frac{1}{4}\bigl(\delta^{\tfrac{1}{4}}+2\,\delta^{\tfrac{5}{4}}\bigr)
+ \delta^{\tfrac{1}{4}}\sqrt{\delta\,\lvert tri(i,j)\rvert}\\
&\;\le\;
\frac{1}{4}\bigl(\delta^{\tfrac{1}{4}}+2\,\delta^{\tfrac{5}{4}}\bigr)
+ \delta^{\tfrac{1}{4}}.
\label{adj_matrix_bound_second_term_est_use_propety_reduction}
\end{aligned}
\end{equation}
By using the Equation \ref{negative_ricii_assumption_conditions_derive_constrain}, we can conclude
\begin{equation}
\begin{aligned}
\frac{1}{\lvert Q_j\rvert}
\sum_{k\in Q_j}(\boldsymbol{\hat A})^2_{ik}
\;\le\;
\delta
\;+\;
\frac{1}{4}\bigl(\delta^{\tfrac{1}{4}}+2\,\delta^{\tfrac{5}{4}}\bigr)
\;+\;
\delta^{\tfrac{1}{4}}
\;\le\;
\,\delta^{\tfrac{1}{4}}.
\label{adj_matrix_bound_Final}
\end{aligned}
\end{equation}
Now, let us revisit the Equation \ref{Llama_1_rewrite}, we have
\begin{equation}
\begin{aligned}
\left\lvert \frac{\partial \boldsymbol{h}_k^{(2)}}{\partial \boldsymbol{h}_i} \right\rvert
\;&\le\;
(\alpha\beta) ( \alpha^{\prime} \beta^{\prime}) \bigl(\boldsymbol{\hat A}\bigr)^2_{ik}\,,
\label{Llama_1_rewrite_revisit}
\end{aligned}
\end{equation}
Finally, we combine the Equation \ref{adj_matrix_bound_Final} and Equation \ref{Llama_1_rewrite_revisit}, we can have the following formulation:
\begin{equation}
\begin{aligned}
\frac{1}{|Q_j|} \sum_{k \in Q_j} \left| \frac{\partial \boldsymbol{h}_k^{(l_0 + 2)}}{\partial \boldsymbol{h}_i^{(l_0)}} \right| < (\alpha \beta) ( \alpha^{\prime} \beta^{\prime}) \delta^{\frac{1}{4}},
\label{negative_ricii_assumption_theory_final}
\end{aligned}
\end{equation}
where we have completed our proof for {\bf Theorem 2}.

\begin{table}[]
        \setlength{\tabcolsep}{1.2pt}
        \caption{Statistics of the datasets for node classification.}
        \label{data_cf}
	\centering
	\begin{tabular}{c||ccccc}
		\toprule
		Dataset    & Instagram & Cora     & Citeseer & Photo      & WikiCS    \\ \midrule
		\# Nodes    & 11,339    & 2,708    & 3,186    & 48,362     & 11,701    \\
		\# Edges    & 144,010   & 5,429    & 4,277    & 500,928    & 216,123   \\
		\# Features & 384       & 1,433    & 384      & 768        & 384       \\
		Domain     & Social    & Academic & Academic & E-commerce & Wikipedia \\ \bottomrule
	\end{tabular}
\end{table}
\begin{table}[]
	\centering
        \caption{Statistics of the datasets for link prediction.}
	\label{data_lp}
	\begin{tabular}{c||cccc}
		\toprule
		Dataset    & Baby       & Sports     & ML-1M     & ML-10M     \\ \midrule
		\# Nodes    & 26,495     & 53,955     & 9,746     & 80,555     \\
		\# Edges    & 160,792    & 296,337    & 1,000,209 & 10,000,054 \\
		\# Features & 384        & 384        & 384       & 384        \\
		Domain     & E-commerce & E-commerce & Movie     & Movie      \\ \bottomrule
	\end{tabular}
\end{table}

\begin{table}[]
    \setlength{\tabcolsep}{1.5pt}
	\centering
        \caption{Statistics of the datasets for graphQA.}
        \label{data_qa}
	\begin{tabular}{c||cccc}
		\toprule
		Dataset     & \# Nodes & \# Edges       & \# Features & Domain       \\ \midrule
		ExplaGraphs & 2,766   & 4.25(Avg)     & 1,024      & Common Sense \\
		WebQSP      & 4,737   & 4,252.37(Avg) & 1,024      & Scene Graph  \\ \bottomrule
	\end{tabular}
\end{table}

\begin{table*}[]
	\centering
        \caption{Performance(\%) comparison of baselines in terms of Recall@5, and Precision@5 on Baby, Sports, MovieLen-1M, and MovieLen-10M datasets. We mark the global best results on each dataset under each metric in boldface and the second best is underlined.}
	\label{res_lp5}
	\begin{tabular}{c|cc|cc|cc|cc}
		\toprule
		Dataset      & \multicolumn{2}{c}{Baby}    & \multicolumn{2}{c}{Sports}    & \multicolumn{2}{c}{MovieLen-1M}     & \multicolumn{2}{c}{MovieLen-10M}    \\ \midrule
		Model        & R@5          & P@5          & R@5           & P@5           & R@5              & P@5              & R@5              & P@5              \\ \midrule
		MF           & 1.86 (0.18)  & 0.41 (0.06)  & 2.92 (0.20)    & 0.58 (0.10)    & 3.96 (0.19)       & 9.25 (0.47)       & 7.04 (0.36)      & 9.21 (0.47)       \\
		MLP          & 1.98 (0.20)  & 0.44 (0.07)  & 3.01 (0.18)    & 0.61 (0.11)    & 4.02 (0.18)       & 9.49 (0.45)       & 7.26 (0.36)      & 9.43 (0.45)       \\ \midrule
		GCN          & 2.06 (0.14)  & 0.49 (0.06)  & 3.05 (0.20)    & 0.62 (0.09)    & 4.15 (0.18)       & 9.83 (0.49)       & 7.45 (0.32)      & 9.81 (0.45)       \\
		GAT          & 2.24 (0.16)  & 0.53 (0.04)  & 3.11 (0.17)    & 0.64 (0.11)    & 4.28 (0.20)       & 10.17 (0.51)      & 7.78 (0.34)      & 10.2 (0.48)       \\
		GIN          & 2.16 (0.14)  & 0.51 (0.07)  & 3.05 (0.19)    & 0.63 (0.09)    & 4.19 (0.17)       & 9.94 (0.43)       & 7.83 (0.30)      & 10.09  (0.39)     \\
		SAGE         & 2.42 (0.11)  & 0.58 (0.06)  & 3.36 (0.18)    & 0.70 (0.08)    & 4.31 (0.20)       & 10.67 (0.46)      & 7.91 (0.30)      & 11.14 (0.39)      \\
		GraphCL      & 2.78 (0.09)  & 0.61 (0.03)  & 3.43 (0.15)    & 0.72 (0.05)    & 4.33 (0.15)       & 10.88 (0.41)      & 8.16 (0.34)      & 11.75  (0.33)     \\ \midrule
		Bert         & 2.01 (0.15)  & 0.47 (0.06)  & 3.09 (0.19)    & 0.64 (0.07)    & 4.17 (0.18)       & 9.85 (0.44)       & 7.56 (0.33)      & 9.81  (0.44)      \\
		RoBERTa      & 2.11 (0.11)  & 0.51 (0.06)  & 3.28 (0.20)    & 0.68 (0.09)    & 4.21 (0.21)       & 9.91 (0.49)       & 7.82 (0.32)      & 10.14 (0.40)     \\ \midrule
		WalkLM       & 2.65 (0.08)  & 0.58 (0.03)  & 3.47 (0.16)    & 0.74 (0.03)    & 4.59 (0.13)       & 11.05 (0.39)      & 8.18 (0.28)      & 11.97  (0.34)     \\
		GraphAdapter & 2.88 (0.10)  & 0.62 (0.04)  & 3.62 (0.14)    & 0.78 (0.02)    & \underline {4.83 (0.17)}       & \underline {11.53 (0.43)}      & OOM              & OOM              \\
		OpenGraph    & 2.96 (0.06)  & 0.61 (0.04)  & 3.51 (0.16)    & 0.73 (0.04)    & 4.43 (0.12)       & 10.94 (0.40)      & 8.33 (0.29)      & 12.16  (0.35)     \\
		TAPE         & 3.09 (0.05)  & 0.64 (0.01)  & 3.86 (0.12)    & 0.80 (0.05)    & 4.69 (0.15)       & 11.32 (0.43)      & 8.36 (0.28)      & 13.08 (0.32)    \\ 
  MuseGraph         & \underline {3.13 (0.07)}  & \underline {0.68 (0.01)}  & \underline {3.89 (0.10)}    & \underline {0.83 (0.03)}    & 4.72 (0.15)       & 11.40 (0.41)      & \underline {8.57 (0.32)}      & \underline {13.43 (0.39)}     \\
  \midrule
		Ours         & \textbf{3.26 (0.05)}  & \textbf{0.72 (0.01)}  & \textbf{4.12 (0.10)}   & \textbf{0.91 (0.02)}   & \textbf{5.20 (0.11)}      & \textbf{12.89 (0.37)}     & \textbf{9.32(0.24)}       & \textbf{14.41(0.29)}      \\ \bottomrule
	\end{tabular}
	
\end{table*}

\subsection{Details for Prompt Design}

In this subsection, we show the detailed information of our prompts used in this work, as shown in Table \ref{tab:prompts}. Note that, our work does not focus on designing various kinds of prompts, but we focus on how to transfer the knowledge from the textual domain to the specific downstream tasks. Therefore, we use the combination of prompts which have been widely used in previous work, like \cite{graphadapter, tape}. Besides, for the link prediction tasks, e.g., Baby, Sports, MovieLen-1M and MovieLen-10M datasets, we only show the textual contents for item nodes, following the most previous works \cite{opengraph, bm3}, where the embedding of user nodes can be obtained by the graph aggregation by item nodes.

\section{Related Work}
In this subsection, we provide introductions to three related areas, i.e. graph representation learning, graph curvature, and large language models with graph.
\subsection{Graph representation learning}Graph representations can be utilized in different applications, including node classification, link prediction, and graph classification. The quality of the graph representation will directly determine the performance of the downstream tasks. Therefore, graph representation learning has become a popular research area. Matrix factorization methods are the early works in graph representation learning. It generates node representations in a framework that learns the proximity measures and SVD decomposition parameters in an end-to-end fashion \cite{DBLP:conf/ijcai/YangSLT17}, \cite{DBLP:conf/kdd/ZhangX0H21}. Graph Neural Network (GNN)-based graph representation learning methods are the second category of graph representation learning methods, which employ GNNs to generate representations. GNN-based methods which are different from traditional methods generalize well to unseen nodes. With the good performance achieved by the classical graph convolutional network \cite{gcn}, it has received more and more attention from researchers and many variants \cite{CGCN},\cite{DSR-GNN} have been developed. Velickovic et al.\cite{GAT} proposed an attention mechanism by learning weights associated with neighboring nodes to distinguish the degree of node influence. He et al. \cite{DBLP:conf/sigir/0001DWLZ020} proposed LightGCN for recommendation tasks which focused on the crucial aspect of neighborhood aggregation. Wang et al.\cite{LND} utilized Laplacian disagreement to jointly assess initial features and output representations to enhance node representations. However, the above methods mainly focus on structural information ignoring the semantic information, which may lead to learn a suboptimal graph representation. 

\subsection{Graph curvature}The geometric concept of curvature has been extended to environments other than smooth manifolds, including on graphs. Ricci curvature is a geometric object which is used to measure the curvature of a Riemannian manifold \cite{jost2008riemannian}. Intuitively, if the Ricci curvature is positive, the curve of the manifold is more spherical. Conversely, if the Ricci curvature is negative, the curve of the manifold is more like a saddle shape. In recent years, more and more researchers have investigated graph curvature, which is a discrete analogue of Ricci curvature \cite{ni2015ricci}, \cite{sia2019ollivier}. Ollivier Ricci curvature is a commonly used curvature measurement method. It has been utilized in the study of complex networks and many works have experimented with its use in GNNs \cite{toppingunderstanding}, \cite{bober2023rewiring}.

\subsection{Large language models with graph}In recent years, large language models (e.g. ChatGPT \cite{chatgpt}, Deepseek \cite{deepseek-llm}, Llama \cite{llama2}) have gained widespread attention due to their remarkable capabilities in various Natuarl Language Processing (NLP) tasks. With the rise open-source LLMs, such as Llama, deepseek, and Vicuna \cite{vicuna2023}, technologies for utilizing pre-trained LLMs to conduct different specific tasks have been proposed, making private LLMs in specific domains possible. Due to the unique capabilities of LLMs, incorporating LLMs into graph tasks presents a promising frontier. A number of studies have emerged at the intersection of graph neural networks and LLMs in recent years. 
Tang et al. \cite{tang2024graphgpt} proposed a framework to align graph domain-specific structural knowledge with the reasoning ability of LLMs to improve the generalization of graph learning.
He et al. \cite{tape} utlized LLMs to generate pseudo labels and explanations as augmented data, and concatenated the enhancements with text feature of PLMs for downstream GNNs. Fang et al. \cite{hituner} distilled hierarchical contextualized representations and incorporated GNNs to structure-related semantics relatedto text-attributed graphs. 
Huang et al. \cite{graphadapter} proposed a residual learning procedure to pre-train the GNN adapter with the LLMs. 
While there have been successful attempts to align LLMs with
graphs, the alignment of text representations with graph representations remains largely unexplored. The alignment process of previous works in LLMs only consider the node information in the graph while ignoring the edge attributes, which leaves the room for improvement in the performance of downstream tasks.
This presents a promising and exciting avenue for future research, and it is precisely this untapped potential that our study aims to explore.

\section{Appendix B for Experiments}
 
\subsection{Datasets}
We use five datasets for node classification \cite{graphadapter}, \cite{engine}, \cite{tape}, \cite{hituner}, four datasets for link prediction \cite{wei2023multi}, \cite{bm3}, and two datasets for graphQA \cite{G-retriever}. 
The datasets used for node classification encompass two citation networks, including Cora and Citeseer. Along with a social network dataset (Instagram), an E-commerce dataset from Amazon (Electronics-Photography, namely Photo) and a Wikipedia-based dataset(WikiCS).
The datasets used for link prediction encompass two E-commerce datasets, including Baby and Sports. Along with two movie datasets from MovieLens (ML-1M and ML-10M).
The datasets used for graphQA encompass ExplaGraphs and WebQSP. ExplaGraphs is a dataset for generative commonsense reasoning and WebQSP is a large-scale multi-hop knowledge graphQA dataset.
The statistics are summarized in Table \ref{data_cf}, Table \ref{data_lp} and Table \ref{data_qa}. In the node classification task, the ratio of nodes used for the train/valid/test stage is 10\%/10\%/80\%, following the previous work \cite{hituner}. In the link prediction task, the ratio of nodes used for the train/valid/test stage is 80\%/10\%/10\%, following the previous work \cite{bm3}. In the graphQA task, the ratio of nodes used for the train/valid/test stage is 60\%/20\%/20\%, which is aligned with previous work \cite{G-retriever}.

\subsection{Implementation details}
In our experiments, we employ LLaMA2 \cite{llama2} as the default LLMs backbone and SAGE as an instance of GNNs. For LLMs, we specify a maximum tokenizer length of $512$ and a batch size of $8$ following the previous work \cite{hituner}. In the case of GNNs, the number of layers is tuned in $\{2, 3\}$.
Adam is adopted as the optimizer. For node classification, the learning rate is tuned in $\{1e-4, 5e-4\}$, and weight decay is set to $0$. The dropout rate is searched in $\{0.5, 0.6\}$. For link prediction, the learning rate is fixed as $1e-3$ and weight decay is searched in $0.1, 0.01$. The dropout rate is chosen from $\{0.3, 0.5\}$. For graphQA, the learning rate is fixed as $1e-5$ and weight decay is set to $0.05$. The dropout rate is set to $0.05$.
For a fair comparison, we run $5$ times and report the mean result and the standard deviation.

Implementations of the baselines are either from open-source projects or the original authors. For the speciﬁc hyperparameters in the baselines, we use the values reported in their original literature. Additionally, we conduct our experiments on Nvidia GeForce RTX 3090 GPUs with 24GB memory.

\begin{figure*}[]
    \centering
    \includegraphics[width=0.8\linewidth]{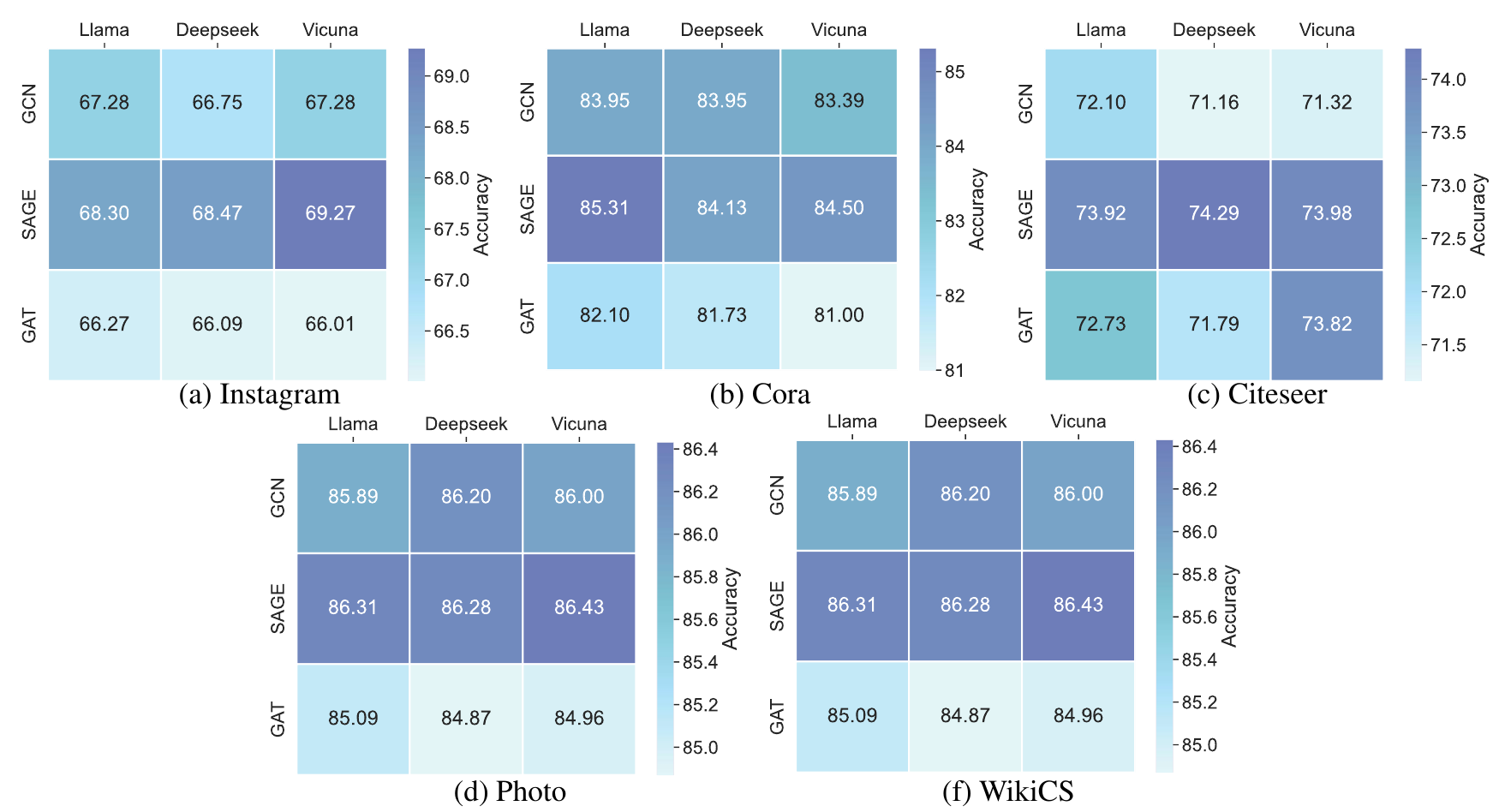}
    \caption{Classification accuracy of CureLLM based on different LLMs and GNNs.
    }
    \label{fig:heat}
\end{figure*}

\begin{table*}[h]
	\centering
        \caption{Ablation study of CureLLM on link prediction task where w/o denotes the removal of a specific component.}
	\label{ablation_lp}
        {
	\begin{tabular}{l|c|c|c|c|c|c|cc}
		\toprule
		                           & \multicolumn{2}{c|}{Baby}     & \multicolumn{2}{c|}{Sports}     & \multicolumn{2}{c|}{MovieLen-1M}    & \multicolumn{2}{c}{MovieLen-10M}    \\ \midrule
		                           & R@10          & P@10          & R@10           & P@10           & R@10             & P@10             & R@10             & P@10             \\ \midrule
		w/o structural information & 4.33          & 0.39          & 5.25           & 0.50           & 6.11             & 9.42             & 10.01            & 9.51             \\
		w/o semantic information   & 4.21          & 0.37          & 5.20           & 0.48           & 6.52             & 9.88             & 11.17            & 10.26            \\
		w/o inference ability      & 4.53          & 0.42          & 5.61           & 0.57           & 7.14             & 9.96             & 12.03            & 10.37            \\
		w/o edge attribute         & \underline{4.98}          & \underline{0.50}          & \underline{6.10}           & \underline{0.62}           & \underline{7.53}             & \underline{10.24}            & \underline{13.55}            & \underline{11.07}            \\ \midrule
		CureLLM                    & \textbf{5.33}          & \textbf{0.59}          & \textbf{6.62}           & \textbf{0.74}           & \textbf{8.60}             & \textbf{11.05}            & \textbf{14.95}            & \textbf{12.18}            \\ \bottomrule
	\end{tabular}
 }
	
\end{table*}

\begin{table*}[h]
\setlength{\tabcolsep}{6pt}
	\centering
        \caption{Ablation study of CureLLM on graph question answering (graphQA) task where w/o denotes the removal of a specific component.}
	\label{ablation_qa}
	\begin{tabular}{c||cccccccccc}
		\toprule
		            & \multicolumn{3}{c}{w/o structural information}         & \multicolumn{3}{c}{w/o inference ability}        & \multicolumn{3}{c}{w/o edge attribute}        & CureLLM \\ \midrule
		ExplaGraphs & \multicolumn{3}{c}{60.03}
  & \multicolumn{3}{c}{54.23}                                                      & \multicolumn{3}{c}{\underline{84.12}}                     & \textbf{87.21}   \\
		WebQSP      & \multicolumn{3}{c}{52.44}
  & \multicolumn{3}{c}{50.26}                                                      & \multicolumn{3}{c}{\underline{69.44}}                     & \textbf{72.58}   \\ \cmidrule{1-4} \cmidrule{6-7} \cmidrule{9-11}
	\end{tabular}
\end{table*}

\subsection{Baselines}
In this subsection, we will provide the descriptions of  the baselines:
\begin{itemize}
    \item \textbf{GCN}: It extracts features by learning relationships between nodes, enabling node representations to capture the topological  structure of the graph.
    \item \textbf{GAT}: It introduces a self-attention mechanism that allows each node to focus on different neighbor nodes.
    \item \textbf{GIN}: It employs a distinct graph encoding method that emphasizes the discrimination of non-isomorphic structures to enhance the representation power of GNNs.
    \item \textbf{SAGE}: It learns an aggregation function, which learns the embedding expression of the target node itself by aggregating the feature information of its neighbors.
    \item \textbf{GraphCL}: It utilizes pre-training of graph models through the application of a self-discriminative contrastive learning task on learned node embeddings.
    \item \textbf{CGCN}: It proposes a new message passing mechanism and integrates it to build an efficient and robust model.
    \item \textbf{DSR-GNN}: It proposes a disparity-induced structural refinement framework, which theoretically dissects the relationship between model capacity and homo/heterophilous ratios to enhance representation learning.
    \item \textbf{BERT}: It captures bidirectional contextual relationships in text through masked language modeling and next-sentence prediction tasks.
    \item \textbf{DeBERTa}: It improves BERT by decoupling content and positional information in attention mechanisms. Its enhanced masked decoder refines contextual modeling.
    \item \textbf{RoBERTa}: It enhances BERT by optimizing the pre-training process: larger data, dynamic masking, and removed next-sentence prediction. 
    \item \textbf{SentenceBERT}: It adapts BERT using siamese and triplet networks and contrastive learning to achieve efficient sentence embeddings.
    \item \textbf{GraphAdapter}: It introduces a graph neural network as an efficient adapter to learn structural information to enrich the semantic information of LLMs.
    \item \textbf{OpenGraph}: It integrates LLMs, a topology-aware graph tokenizer and a scalable graph transformer into one framwork which captures universal and transferable structural patterns across multiple domains.
    \item \textbf{TAPE}: It uses LLMs to generate  pseudo labels and explanations as augmented data, and concatenates the enhancements with original text for PLMs.
    \item \textbf{WalkLM}: It performs attributed random walks on the graph to compose roughly meaningful textual sequences then extracts embedding vectors from the LMs.
    \item \textbf{ENGINE}: It combines LLMs and GNNs by a tunable side structure to adopt message passing at  each layer of LLMs to integrate structural information.
    \item \textbf{GraphToken}: It is a parameter-efficient encoder for structured data inclusion in LLMs which is used for various reasoning tasks.
    \item \textbf{G-Retriever}: It presents a retrieval approach for general textual graph tasks, which significantly enhances scalability and efficiency.
    \item \textbf{STAG}: It employs soft assignment and KL divergence guided quantization to address the unique challenges of graph data, which lacks natural tokenization structures.
    \item \textbf{MuseGraph}: It seamlessly integrates the strengths
of GNNs and LLMs into one foundation model for graph
mining across tasks and datasets.
    
\end{itemize}

\subsection{Additional Performance Comparison}


We compare the link prediction results of our \methodbase to those of the baseline models. Table \ref{res_lp5} shows the ${Recall@5}$ and ${Precision@5}$ on Baby, Sports, MovieLen-1M (ML-1M) and MovieLen-10M (ML-10M) datasets.
From Table \ref{res_lp5}, we have the following observations: In general, \methodbase outperforms all $13$ baselines across all evaluation metrics on all datasets. It gains a substantial lead over the second-best models, TAPE and GraphAdapter, with performance improvements ranging from $5.5\%$ in ${Recall@5}$ on Baby to a significant $11.5\%$ on MovieLen-10M. Baselines that combine structural and semantic information generally perform better than baselines that use one type of information alone. Compared LLM-enhanced methods, our \methodbase achieves the highest performance across all datasets, which also indicates the effectiveness of edge attributes.

\subsection{Ablation Study}
We evaluate all components by removing each item from \methodbase in turn: (a) \textit{w/o} structural information: replacing GNN with MLP; (b) \textit{w/o} semantic information: removing textual embedding; (c) \textit{w/o} inference ability: replacing Llama2 with DeBERTa; (d) \textit{w/o} edge attribute: removing curvature-based alignment. 
Experimental results for node classification, link prediction, and graph question answering (GraphQA) are listed in Tables \ref{ablation_cf}, \ref{ablation_lp}, and \ref{ablation_qa}, respectively. Notably, unlike other tasks, semantic information is crucial in graph question answering. As shown in the tables, removing any component leads to a performance degradation. Table \ref{ablation_cf} shows that removing any component leads to a performance degradation in node classification, demonstrating the indispensability of all four components to the overall model performance.
Table \ref{ablation_lp} shows that the performance degradation caused by removing structural and semantic information is significantly greater than that caused by removing other components, indicating that both structural and semantic information are beneficial for link prediction. 
From Table \ref{ablation_qa}, we can clearly observe that the performance degradation of (a) \textit{w/o} structural information and (c) \textit{w/o} inference ability is particularly significant, indicating that both structural information and rich semantic information play a pivotal role in the graphQA task. In addition, the performance degradation of (d) \textit{w/o} edge attribute means that mining edge attributes for aligning the graph and textual information is capable of learning a more effective representation.

\subsection{Scalability Study}
We conduct a scalability study focusing on different combinations of LLMs (LLama2-7B \cite{llama2}, Deepseek-chat-7B \cite{deepseek-llm}, Vicuna1.5-7B \cite{vicuna2023}) and GNNs (GCN \cite{gcn}, SAGE \cite{sage}, GAT \cite{GAT}). As depicted in Figure \ref{fig:heat}, in each subplot, there is little change in color when viewed horizontally. This shows the strong compatibility of the proposed method for LLM selection. Furthermore, in each subplot, the change in color is more obvious when viewed vertically. This implies that the learning ability of structural information is key to the model performance. Overall, there is very slight change in performance for different combinations, indicating that CureLLM is a scalable framework proficient in adapting to diverse architectures.

\end{document}